\documentclass[useAMS]{mn2e}

\usepackage[dvips]{graphicx}
\usepackage{verbatim}
\usepackage{lscape}

\title[Optical and radio polarization of AGNs II]{Search for correlations between the optical and radio polarization of AGNs II: VLBA polarization data at 12+15+22+24+43~GHz}

\author[Algaba et.\ al.]{J. C. Algaba$^{1}$\thanks{algaba@asiaa.sinica.edu.tw}, D. C. Gabuzda$^{2}$ and P. S. Smith$^{3}$\\
$^{1}$Academia Sinica, Institute of Astronomy and Astrophysics, P.O. Box 23-141, Taipei 106, Taiwan R. O. C\\
$^{2}$Department of Physics, University College Cork, Cork, Ireland\\
$^{3}$Steward Observatory, The University of Arizona, Tucson, AZ, USA}
\begin{document}

\date{Accepted ---. Received ---; in original form ---}

\pagerange{\pageref{firstpage}--\pageref{lastpage}} \pubyear{2011}

\maketitle

\label{firstpage}

\begin{abstract}

Previous research showed that most BL Lac objects and some quasars have aligned VLBI-core and optical polarizations, although some of the AGNs also showed no obvious relationship between their VLBI-core and optical polarization angles. This may indicate that some AGNs have co-spatial regions of optical and radio emission, while others do not. However, another possibility is that some of the VLBI cores had Faraday rotations of the order of several tens of thousand of rad/m$^2$, which were not properly fit using the three-frequency data due to $n\pi$ ambiguities in the observed polarization angles, leading to incorrect subtraction of the effects of the core Faraday rotation, and so incorrect intrinsic radio polarization angles $\chi_0$. With this is mind, we obtained additional 12+15+22+24+43~GHz plus optical observations for 8 of 40 AGNs previously considered, enabling improved sampling of Faraday effects. Our results indicate that, although some VLBI radio cores have comparatively high rotation measures, this alone cannot explain the misalignments found between the radio core and optical VLBI polarization angles, $\Delta\chi = |\chi_{opt} - \chi_0|$. Comparison between $\Delta\chi$ and (i) the orientation of $\chi_0$ relative to the jet direction, (ii) the degree of polarization of the core, (iii) a depolarization factor, (iv) the core rotation measures and (v) the core magnetic fields 1~pc from the jet base do not yield evidence for any correlations between these properties. 
There is,however, some evidence that the maximum observed $\Delta\chi$ tends to decrease as the core-region magnetic field increases, suggesting that large misalignments in $\Delta\chi$ could be associated in part with relatively low core magnetic fields.
Thus, although the overall distribution of $\Delta\chi$ for all 40 sources in our sample does show a significant peak at $\Delta\chi\sim 0$, it remains unclear what distinguishes these AGN cores from those showing appreciable misalignment between optical and VLBI-core polarization position angles.
\end{abstract}



\section{Introduction}
In general, non-thermal synchrotron emission associated with the relativistic jets of radio-loud AGNs dominates the UV--radio continua of radio--loud active galactic nuclei (AGNs). Even so, little correlation between the observed emission in widely spaced wavebands has usually been expected, even if genuinely simultaneous observations are compared. This is in part because early attempts to search for correlations using integrated single--dish measurements were largely unsuccessful (e.g., Rudnick 1978). In addition, optical variations typically occur more rapidly that radio variations, suggesting that the optical emission is generated in a much smaller volume than the radio emission. In this context, it was considered natural for there to be little or no correlation between the optical and radio emission in AGNs.

It has only been relatively recently that correlations between properties at different wavelengths have been found, with   
several investigations finding evidence for the polarization properties of sources at different wavebands to be correlated. For example, Lister \& Smith (2000) carried out a joint analysis of VLBA polarization data at 22 and 43~GHz and optical polarization data. Despite the fact that the optical and VLBA measurements were obtained about a month apart, they found evidence that the optical polarization follows the radio structure and tends to lie along the VLBI jet direction, with the degree of optical polarization correlated with the core polarization and core luminosity at 43~GHz.

More recently, Gabuzda et al. (2006) carried out a simultaneous analysis of optical polarization data and 15+22+43~GHz VLBI 
polarization data for 11 BL Lac objects and 3C279. Their results showed that, after correction for Faraday rotation, nearly all the VLBI core polarizations were aligned with the corresponding optical polarizations within 20$^\circ$. Similar results were obtained by Jorstad et. al (2007) at even shorter radio wavelengths. Also, a co-rotation of the 7mm and optical polarization angle of 0420--014 through $\sim80^{\circ}$ over about 10 days was found by D'Arcangelo et al (2007).

However, the optical and 15+22+43~GHz observations analyzed by Algaba et. al (2010) (hereafter, Paper I) gave rise to unexpected results. Based on the work of Gabuzda et. al (2006), the initial expectation was a correlation between the optical and Faraday-corrected VLBA radio-core polarization angles, with the difference between these two observed angles, $\Delta\chi$, often being close to zero. As seen in Paper I, this is not the case: not only do quasars in general not show this trend, but neither do all BL Lacs, in contrast with the striking results found by Gabuzda et. al (2006).

In Paper I we tried to identify some possible reasons for the lack of correlation in some sources. In particular, we searched for relations between $\Delta\chi$ and the degrees of polarization at the observed radio frequencies, and between $\Delta\chi$ and the difference between the intrinsic radio-core polarization angle and jet direction. No signs of any relation were found in either case.

Paper I discusses possible reasons for this such as the existence of internal or high external Faraday Rotation, but, unfortunately, the data were not able to unambiguously discriminate between these possibilities. Although three frequencies are in principle adequate to fit for the Faraday rotation and find the intrinsic polarization angle, a variety of uncertainties and ambiguities can arise due to factors such as the Faraday rotation effects mentioned above, uncertainty about optical depth transitions, and beam depolarization. 
We investigate these issues here using five-frequency Very Long Baseline Array polarization data for a subset of eight of the 40 AGNs.

\section{Observations}

\subsection{VLBA}

We obtained a new set of polarization observations on the NRAO Very Long Baseline Array (VLBA) at 12.039, 15.383, 21.775, 23.998 and 43.135~GHz, each with two intermediate-frequency bands with a bandwidth of 8~MHz and an aggregate bitrate of 128~MB/s. These polarization observations were carried out in a 24-hour session on November 2, 2008 (hereafter, epoch D). In all cases, the sources were observed in a ``snapshot'' mode, with 8--10 several-minute scans for each object at each frequency spread in time. The resulting coverage in the $u$--$v$ plane was quite uniform. The data reduction and imaging for the radio data were done with the NRAO Astronomical Image Processing System (AIPS) using standard techniques. The reference antenna used was Los Alamos. Simultaneous solutions for the instrumental polarizations and source polarizations for the compact source 1954+513 were derived using the AIPS task LPCAL.

We calibrated the electric vector position angles (EVPAs) using the phases of the VLBA polarization D-terms (G\'omez et. al 1992). Comparison of the D-terms against a set of tabulated values (previously calibrated by other means) is a reliable method for calibrating the absolute L--R phase offset of the reference antenna in VLBA observations. We were able to use two sets of data at the same or very similar frequencies, for which the EVPA calibration had been performed using integrated polarization measurements in the standard way (O'Sullivan \& Gabuzda 2009; Reichstein \& Gabuzda 2011), making it possible to apply this method to the data. These other sets of data were obtained on July 2, 2006 (12.9, 15.4, 22.2, 43.1~GHz) and September 27, 2007 (12.9, 15.4~GHz). 

To ensure the reliability of this method, we performed a series of additional checks. First, we compared the independent results given by the two comparison sets of D-terms for consistency. Second, we compared the new images with previous data (Algaba 2010) for the 8 AGNs in common. Third, for some frequencies, we were also able to compare the resulting polarization angles with images from the MOJAVE\footnote{http://www.physics.purdue.edu/MOJAVE/} and Boston University\footnote{http://www.bu.edu/blazars/VLBAproject.html} databases, as well as integrated measurements from the University of Michigan\footnote{http://www.astro.lsa.umich.edu/obs/radiotel/umrao.php and M. Aller, personal communication} database. Fourth, we obtained the EVPAs of several sources in the optically thin jet regions and checked for consistency with modest Faraday rotation. We estimate the overall error in the EVPA calibration to be about $4^{\circ}$ for all frequencies. We summarize the results of our EVPA calibrations in Table~\ref{EVPAcorr}.

\begin{table}\begin{center}
\caption{EVPA Corrections}
\begin{tabular}{ c c c c c }
\hline
12GHz & 15 GHz & 22 GHz &24 GHz & 43 GHz \\
\ [1]&[2]&[3]&[4]&[5]\\
\hline
$66\pm4$ & $-63\pm4$ & $45\pm4$ & $-85\pm4$ & $-20\pm4$\\
\hline
\label{EVPAcorr}
\end{tabular}\end{center}
\end{table}

We found a systematic flux excess for all sources at 24~GHz, presumably due to the application of a somewhat incorrect gain during the amplitude calibration of these data. Checking the total and polarized fluxes at several points in several sources and comparing with the results for the corresponding 12, 15, 22 and 43~GHz images indicated that the total and polarized fluxes at 24~GHz were both consistently about 15--20\% too high. Although this will affect measurements involving the flux density at 24~GHz, it will not affect the inferred degrees of polarization (m) or $\chi$ values, since these are determined using ratios. 
We thus include the 24~GHz degrees of polarization and $\chi$ values in our analysis, but the 24~GHz measurements are not used in the calculations of the spectral indices. 

After the initial construction of the total intensity maps, we used the final self-calibrated visibility data to make maps of the 
distributions of the Stokes parameters $Q$ and $U$ at each frequency. Those 
maps were used to construct the distribution of polarized flux ($p=
\sqrt{Q^2+U^2}$) and polarization angle ($\chi=\frac{1}{2}\arctan\frac{U}{Q}$), 
using the AIPS task COMB. We obtained the $\chi$ values at specified locations 
and their uncertainties using the AIPS facility IMSTAT applied to a $3\times3$ pixel 
region (i.e., $0.15\times0.15$mas$^2$, as each pixel corresponds to 0.05 mas) 
surrounding the location of interest, with an estimated EVPA calibration uncertainty
of $4^{\circ}$ added in quadrature.

\subsection{Optical}

Nightly optical polarization observations spanned the VLBA observing run. Data were acquired using the SPOL spectropolarimeter (Schmidt et. al 1992a) at the 2.3m Bok telescope of Steward Observatory. On various nights, the instrument was configured for spectropolarimetry using a 600 line/mm diffraction grating. Data acquisition and reduction closely follow those described by Smith et. al (2003) and Gabuzda et. al (2006). The spectropolarimetric observations are averaged over the R filter bandpass for direct comparison to the imaging polarimetry. The interstellar polarization standard stars BD+59$ö{\circ}$389, BD+64$ö{\circ}$106, HD 236633, HD 245310, and Hiltner 960 (Schmidt et al. 1992b) were observed to calibrate $\chi_{opt}$.

Optical measurements for 1633+382 are publicly available for the Fermi program\footnote{http://james.as.arizona.edu/$\sim$psmith/Fermi}. For some of the sources observed in radio, scans ranged all over the 24 hour observation. This was the case for 0256+075, 0420--014, 0906+430 and 1633+382. In these cases, an average of the optical values corresponding to days November 1 and 2, 2008 was taken. Both the optical and radio polarization measurements were corrected for statistical biases (Wardle \& Kronberg 1974).

\section{Results}

\subsection{Intensity and Polarization of the Cores}

The total intensity, polarization and spectral-index maps for all eight sources observed are presented by Algaba (2010) and are also available by contacting  J.\ C.\ Algaba. 
Total and polarized intensity properties of the VLBI core components are shown in Table~2. Column [1] shows the name of the source, column [2] the frequency, column [3] the peak total intensity and column [4] the degree of polarization obtained by applying IMSTAT in AIPS over a $3\times3$ pixel area surrounding the intensity peak. 

We find a systematic frequency dependence in 7 out of 8 cases, where the optical $m$ is higher than the measured $m$ at radio frequencies. However, in the radio band, there is no universal behaviour for the frequency dependence of the core $m$. There is some tendency for $m$ to increase with increasing radio--frequency for 0133+476, 0745+241, and 0906+430, while 0256+075 shows the opposite tendency, consistent with the inferred core spectral indices. There is no obvious systematic radio--frequency dependence of $m$ for the remaining four objects. 

All the sources have a bright core with a jet consisting of several components, with the innermost jet component not well resolved from the core at our lowest observed frequencies. The polarization angle varies with frequency in the core, suggesting Faraday rotation (discussed further below), while the polarization angles are normally fairly well aligned in the jet, indicating only modest Faraday rotation. This agrees with our expectation that both the magnetic field and the electron density decrease with distance from the core.

\begin{table}
\begin{center}
\caption{Total Intensity and Degree of Polarization}
\begin{tabular}{ c c c c }
\hline\hline
Source  & Frequency (GHz) & Ipeak (mJy) & m (\%)\\
\ [1]&[2]&[3]&[4]\\
\hline
0133+476&		12	&$	3180	\pm	20	$&$	1.1	\pm	0.2	$\\
		&	15 	&$	3060	\pm	20	$&$	1.5	\pm	0.1	$\\
		&	22 	&$	2990	\pm	20	$&$	2.6	\pm	0.1	$\\
		&	24 	&$	-				$&$	2.5	\pm	0.1	$\\
		&	43 	&$	2370	\pm	20	$&$	2.5	\pm	0.1	$\\
	&	Optical 	&$	-		$&$	3.8	\pm	0.1	$\\
\hline
0256+075	&	12 	&$	316	\pm	6	$&$	5.4	\pm	0.3	$\\
		&	15 	&$	328	\pm	4	$&$	4.6	\pm	0.2	$\\
		&	22 	&$	378	\pm	4	$&$	2.6	\pm	0.1	$\\
		&	24 	&$	-			$&$	3.1	\pm	0.1	$\\
		&	43 	&$	389	\pm	2	$&$	1.0	\pm	0.1 $\\
	&	Optical 	&$	-		$&$	20.5	\pm	0.1	$\\
\hline
0420-014	&	12 	&$	3710	\pm	20	$&$	2.2	\pm	0.1	$\\
		&	15 	&$	3580	\pm	40	$&$	2.4	\pm	0.1	$\\
		&	22 	&$	3240	\pm	20	$&$	2.4	\pm	0.1	$\\
		&	24 	&$	-				$&$	2.7	\pm	0.1	$\\
		&	43 	&$	2270	\pm	40	$&$	1.5	\pm	0.1	$\\
	&	Optical 	&$	-		$&$	3.9	\pm	0.1	$\\
\hline
0745+241	&	12 	&$	1390	\pm	20	$&$	1.3	\pm	0.1	$\\
		&	15 	&$	1330	\pm	10	$&$	1.8	\pm	0.1	$\\
		&	22 	&$	1310	\pm	9	$&$	2.5	\pm	0.1	$\\
		&	24 	&$	-				$&$	2.6	\pm	0.1	$\\
		&	43 	&$	1067	\pm	8	$&$	3.5	\pm	0.1	$\\
	&	Optical 	&$	-		$&$	6.7	\pm	0.1	$\\
\hline
0906+430	&	12 	&$	694	\pm	6	$&$	0.2	\pm	0.1	$\\
		&	15 	&$	622	\pm	4	$&$	0.3	\pm	0.1	$\\
		&	22 	&$	590	\pm	4	$&$	1.9	\pm	0.1	$\\
		&	24 	&$	-			$&$	2.1	\pm	0.1	$\\
		&	43 	&$	439	\pm	4	$&$	1.2	\pm	0.1	$\\
	&	Optical 	&$	-		$&$	4.5	\pm	0.2	$\\
\hline
1633+382	&	12 	&$	2110	\pm	20	$&$	1.2	\pm	0.1	$\\
		&	15 	&$	1990	\pm	10	$&$	0.4	\pm	0.1	$\\
		&	22 	&$	2020	\pm	10	$&$	0.2	\pm	0.1	$\\
		&	24 	&$	-				$&$	0.5	\pm	0.1	$\\
		&	43 	&$	1920	\pm	20	$&$	1.3	\pm	0.1	$\\
	&	Optical 	&$	0.33\pm4		$&$	3.0	\pm	0.1	$\\
\hline
1823+568	&	12 	&$	1180	\pm	10	$&$	7.4	\pm	0.3	$\\
		&	15 	&$	1131	\pm	7	$&$	7.6	\pm	0.1	$\\
		&	22 	&$	1089	\pm	7	$&$	9.2	\pm	0.1	$\\
		&	24 	&$	-			$&$	8.9	\pm	0.1	$\\
		&	43 	&$	887	\pm	8	$&$	5.7	\pm	0.1	$\\
	&	Optical 	&$	-		$&$	22.9	\pm	0.1	$\\
\hline
1954+413	&	12 	&$	1200	\pm	10	$&$	1.4	\pm	0.1	$\\
		&	15 	&$	1161	\pm	9	$&$	1.8	\pm	0.1	$\\
		&	22 	&$	1131	\pm	8	$&$	1.1	\pm	0.1	$\\
		&	24 	&$	-				$&$	1.4	\pm	0.1	$\\
		&	43 	&$	709		\pm	8	$&$	0.7	\pm	0.1	$\\
	&	Optical 	&$	-		$&$	1.1	\pm	0.2	$\\

\hline
\end{tabular}
\end{center}
\label{tab:iandp}
\end{table}

\subsection{Core Spectral Indices}

It is of interest to derive the spectral indices for the cores, to determine whether the core regions are predominantly optically thin or optically thick.
Unfortunately, during the process of VLBI calibration and imaging, information about the absolute position of the source is lost. In practice this means that, prior to any comparison of maps obtained at two different frequencies, we must align the maps. This can be done in several ways: using astrometric phase--referenced observations to compare the position of the target source relative to a reference point--like source (see e.g., Pradel et. al 2006) or cross-correlating the optically thin regions of the jet at pairs of frequencies (see e.g., Croke \& Gabuzda 2008). We use here the latter approach.

For the construction of the spectral-index maps, we made versions of the final 
images using the final calibrated data but convolving the images at all frequencies with the 
beam obtained for the 15~GHz map. This beam was chosen as a compromise 
between the lower (12~GHz) and higher (22, 24, 43~GHz) resolution maps; although we 
are super-resolving slightly at 12~GHz, this is only by a modest amount. We 
then used the task COMB to obtain the spectral-index maps after aligning the 
images using the program developed by Croke and Gabuzda (2008). 

Table~3 contains the core shifts at each frequency relative to 43~GHz in milliarcseconds. In each case, the derived core shifts are roughly opposite to the innermost jet directions, as expected (the position of the ``core'' moves down the jet at lower frequencies). Note that we have included the 24~GHz data in our core-shift analysis: although the absolute scale of our 24~GHz images is inaccurate, the form of the image structure is correct, so that a correlation of the 24~GHz and 43~GHz images yields a correct relative shift between them.
The shifts between 22 and 24~GHz are small compared with the shifts between other pairs of frequencies, as would be expected, and the shifts obtained for the 24-GHz images are consistent with the other relative shifts obtained. Of course, we are not able to calculate spectral indices using the 24-GHz images due to the flux-scale problem described in \S2.1.

The core spectral indices (defined as $S\propto \nu^{+\alpha}$) derived after aligning the maps at pairs of frequencies are shown in Table~4. The frequencies of the two maps taken to obtain the data are shown in the second column. The spectral index for the core is shown in the third column. Most of the source cores seem to be appreciably optically thin; i.e., their spectral indices are negative. 
The quasar 1633+382 shows a somewhat flatter spectrum, and the BL Lac 0256+075 an inverted spectrum. 

\begin{table}
\begin{center}
\caption{Core Shifts}
\begin{tabular}{ c c c c }
\hline\hline
Source  & Frequency Pair            & \multicolumn{2}{|c|}{Core Shift} \\
        & Used (GHz)		& Magnitude (mas) & Direction ($^{\circ}$) \\
\ [1]&[2]&[3]&[4]\\
\hline

0133+476&	12--43	&$	0.10	\pm0.04$& $135\pm8^{\circ}$ \\
	&	15--43	&$	0.07	\pm0.04$& $135\pm8^{\circ}$\\
	&	22--43	&$	0.04	\pm0.04$& $135\pm8^{\circ}$\\
	&	24--43	&$	0.04	\pm0.04$& $135\pm8^{\circ}$\\
\hline
0256+075&	12--43	&$	0.16	\pm0.04$& $26\pm5^{\circ}$\\
	&	15--43	&$	0.11	\pm0.04$& $26\pm5^{\circ}$\\
	&	22--43	&$	0.06	\pm0.04$& $26\pm5^{\circ}$\\
	&	24--43	&$	0.06	\pm0.04$& $26\pm5^{\circ}$\\
\hline
0420-014&	12--43	&$	0.18	\pm0.04$& $0\pm4^{\circ}$\\
	&	15--43	&$	0.14	\pm0.04$& $0\pm4^{\circ}$\\
	&	22--43	&$	0.08	\pm0.04$& $0\pm8^{\circ}$\\
	&	24--43	&$	0.05	\pm0.04$& $0\pm4^{\circ}$\\
\hline
0745+241&	12--43	&$	0.08	\pm0.04$& $180\pm8^{\circ}$\\
	&	15--43	&$	0.05	\pm0.04$& $116\pm10^{\circ}$\\
	&	22--43	&$	0.02	\pm0.04$& $90\pm8^{\circ}$\\
	&	24--43	&$	0.02	\pm0.04$& $90\pm8^{\circ}$\\
\hline
0906+430&	12--43	&$	0.14	\pm0.04$& $-45\pm8^{\circ}$\\
	&	15--43	&$	0.11	\pm0.04$& $-45\pm8^{\circ}$\\
	&	22--43	&$	0.08	\pm0.04$& $-45\pm16^{\circ}$\\
	&	24--43	&$	0.05	\pm0.04$& $-45\pm8^{\circ}$\\
\hline
1633+382&	12--43	&$	0.10	\pm0.04$& $135\pm8^{\circ}$\\
	&	15--43	&$	0.07	\pm0.04$& $135\pm8^{\circ}$\\
	&	22--43	&$	0.04	\pm0.04$& $135\pm16^{\circ}$\\
	&	24--43	&$	0.04	\pm0.04$& $135\pm8^{\circ}$\\
\hline
1823+568&	12--43	&$	0.16	\pm0.04$& $26\pm5^{\circ}$\\
	&	15--43	&$	0.10	\pm0.04$& $26\pm10^{\circ}$\\
	&	22--43	&$	0.05	\pm0.04$& $26\pm20^{\circ}$\\
	&	24--43	&$	0.03	\pm0.04$& $26\pm10^{\circ}$\\
\hline
1954+413&	12--43	&$	0.13	\pm0.04$& $135\pm8^{\circ}$\\
	&	15--43	&$	0.10	\pm0.04$& $135\pm8^{\circ}$\\
	&	22--43	&$	0.06	\pm0.04$& $135\pm16^{\circ}$\\
	&	22--43	&$	0.02	\pm0.04$& $135\pm8^{\circ}$\\
\hline
\end{tabular}
\end{center}
\label{tab:coreshift}
\end{table}

\begin{table}
\begin{center}
\caption{Core Spectral Indices}
\begin{tabular}{ c c c }
\hline\hline
Source  & Frequency Pair        & Spectral Index \\
        & Used (GHz)		& 		 \\
\ [1]&[2]&[3]\\
\hline

0133+476&	12--15	&$	-0.18	\pm0.08	$\\
	&	15--22	&$	-0.07	\pm0.08	$\\
	&	22--43	&$	-0.34	\pm0.10	$\\
\hline
0256+075&	12--15	&$	 +0.40	\pm0.20	$\\
	&	15--22	&$	 +0.53	\pm0.14	$\\
	&	22--43	&$	 +0.16	\pm0.20	$\\
\hline
0420-014&	12--15	&$	-0.20	\pm0.10	$\\
	&	15--22	&$	-0.27	\pm0.11	$\\
	&	22--43	&$	-0.64	\pm0.08	$\\
\hline
0745+241&	12--15	&$	-0.19	\pm0.12	$\\
	&	15--22	&$	-0.04	\pm0.10	$\\
	&	22--43	&$	-0.29	\pm0.08	$\\
\hline
0906+430&	12--15	&$	-0.46	\pm0.10	$\\
	&	15--22	&$	-0.14	\pm0.09	$\\
	&	22--43	&$	-0.44	\pm0.09	$\\
\hline
1633+382&	12--15	&$	-0.25	\pm0.09	$\\
	&	15--22	&$	 +0.04	\pm0.08	$\\
	&	22--43	&$	-0.07	\pm0.07	$\\
\hline
1823+568&	12--15	&$	-0.20	\pm0.11	$\\
	&	15--22	&$	-0.08	\pm0.08	$\\
	&	22--43	&$	-0.30	\pm0.05	$\\
\hline
1954+413&	12--15	&$	-0.15	\pm0.10	$\\
	&	15--22	&$	-0.09	\pm0.10	$\\
	&	22--43	&$	-0.68	\pm0.07	$\\
\hline
\end{tabular}
\end{center}
\label{tab:spix}
\end{table}

\section{Search for Correlations in $\chi$}

We obtained the $\chi$ values of the core regions by applying the AIPS task IMSTAT to corresponding $3\times 3$-pixel areas at all 5 frequencies. For consistency, we checked other nearby regions for each source to make sure the $\chi$ values we were obtaining were stable around the central measured region. We then fitted the obtained $\chi$ values with a linear $\lambda^2$ law to obtain the Faraday rotation measure (RM) and the intrinsic $\chi_0$: $\chi = \chi_0 + \textrm{RM}\lambda^2$. In contrast to our previous observations (Paper I), we now have five frequencies, two of them (22 and 24~GHz) close enough so that we should be able to unambiguously determine if $n\pi$ rotations are needed for the lower--frequency $\chi$ values. 

\begin{figure*}
\begin{center}
\includegraphics[width=6.8cm]{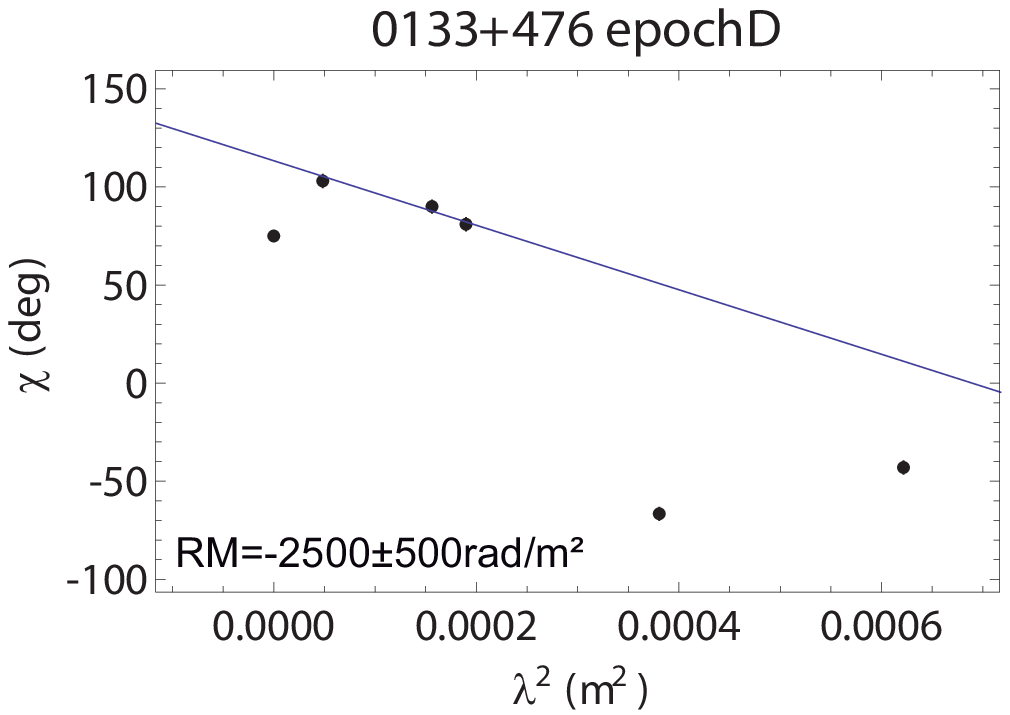}\hspace{0.8cm}
\includegraphics[width=6.8cm]{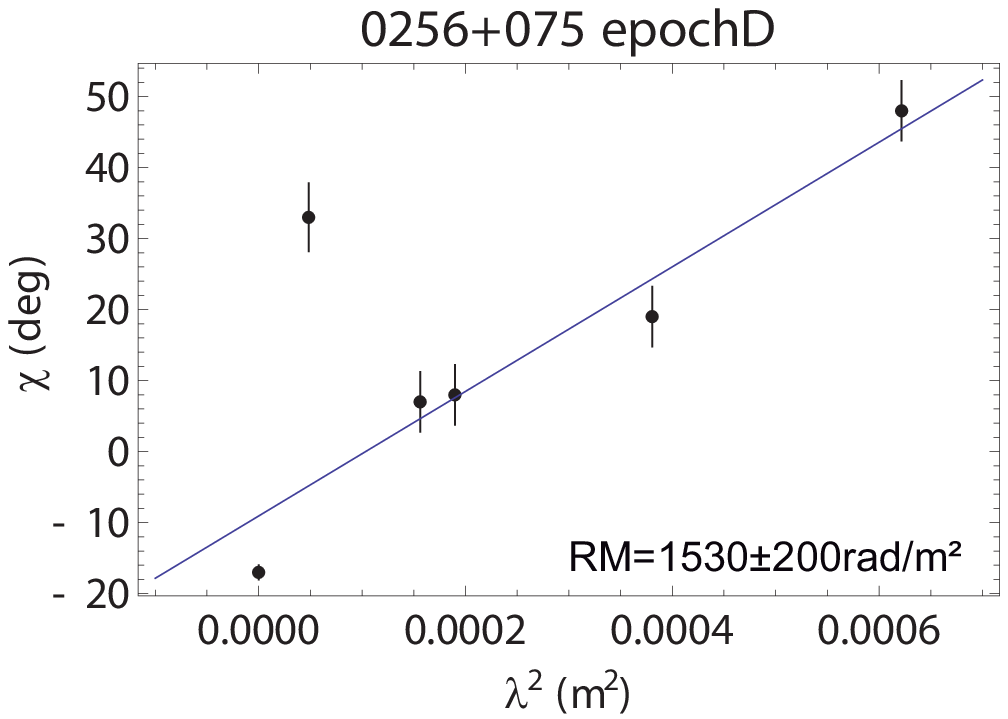}\\

\includegraphics[width=6.8cm]{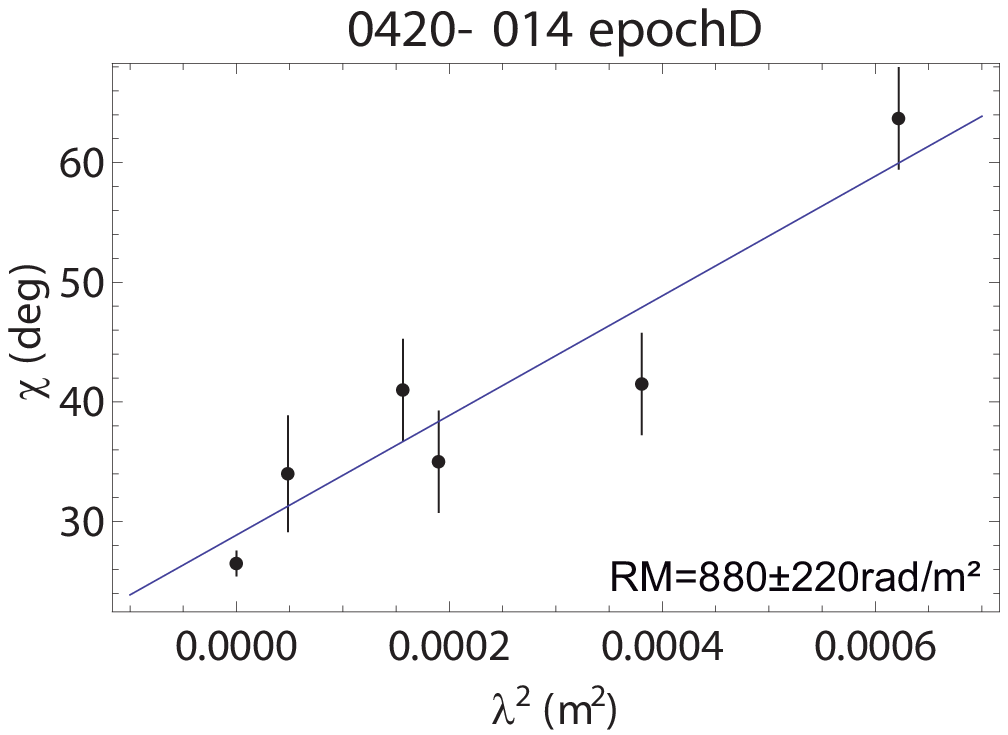}\hspace{0.8cm}
\includegraphics[width=6.8cm]{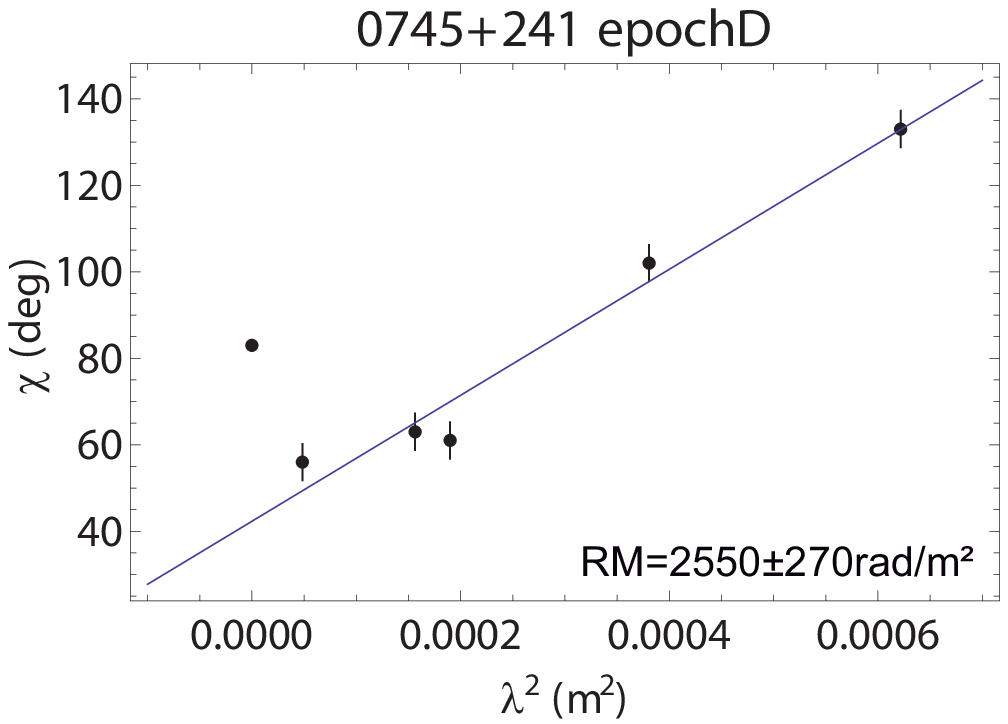}\\

\includegraphics[width=6.8cm]{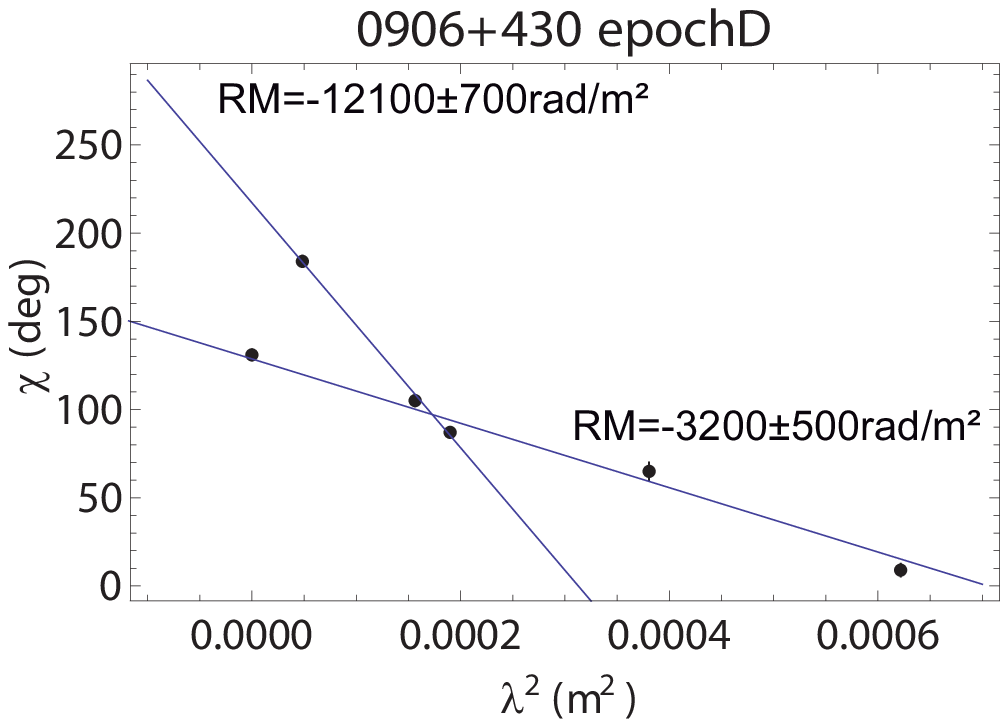}\hspace{0.8cm}
\includegraphics[width=6.8cm]{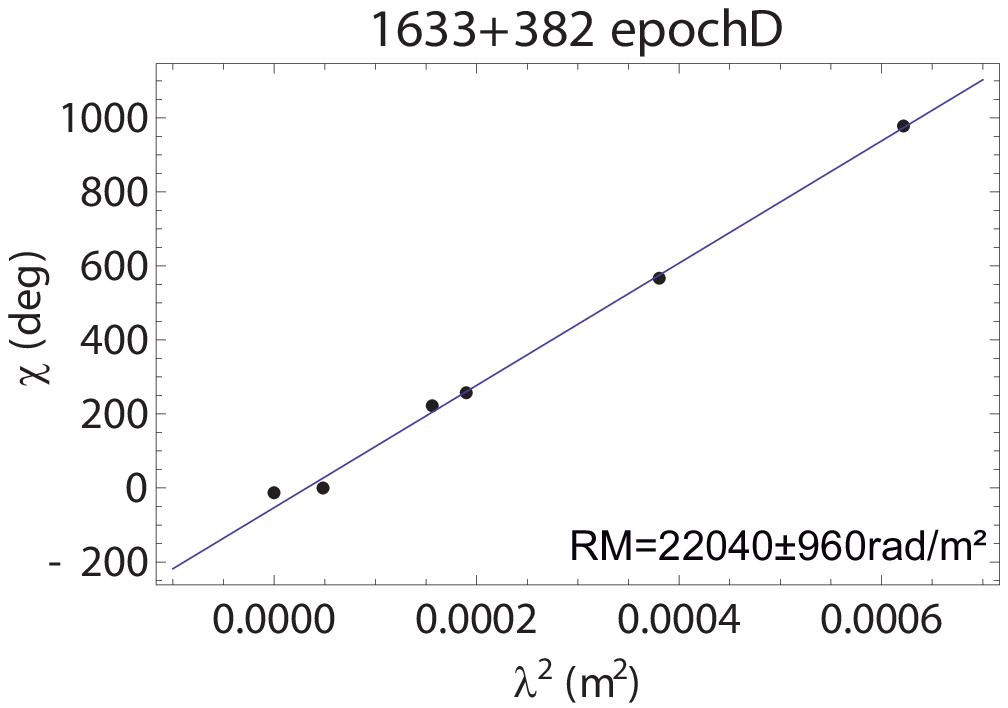}\\

\includegraphics[width=6.8cm]{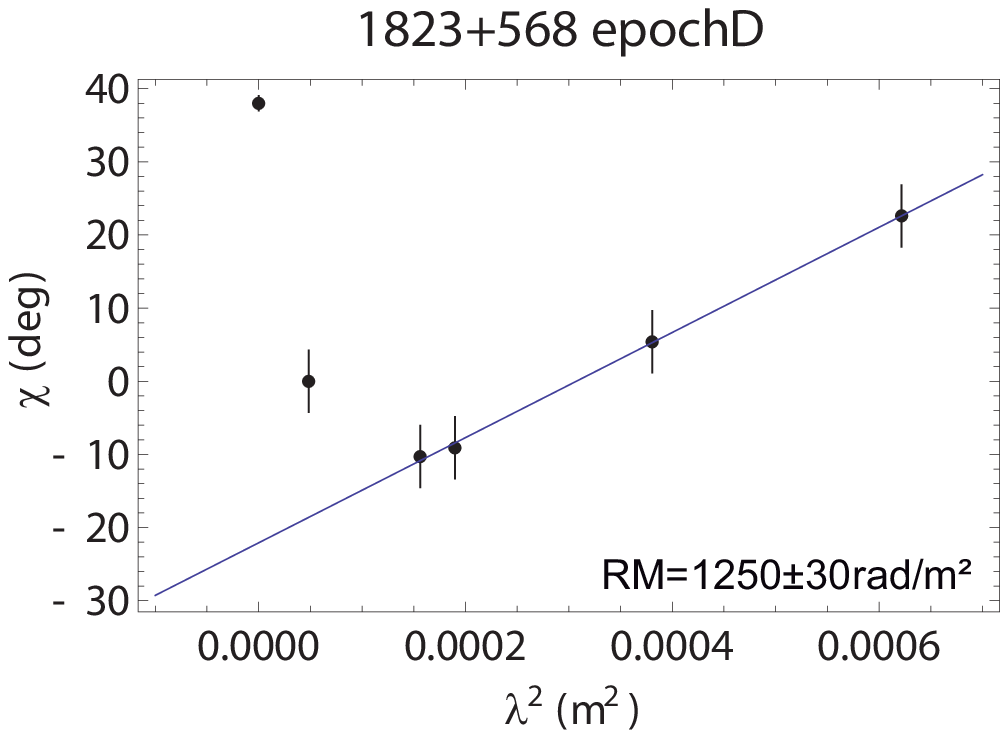}\hspace{0.8cm}
\includegraphics[width=6.8cm]{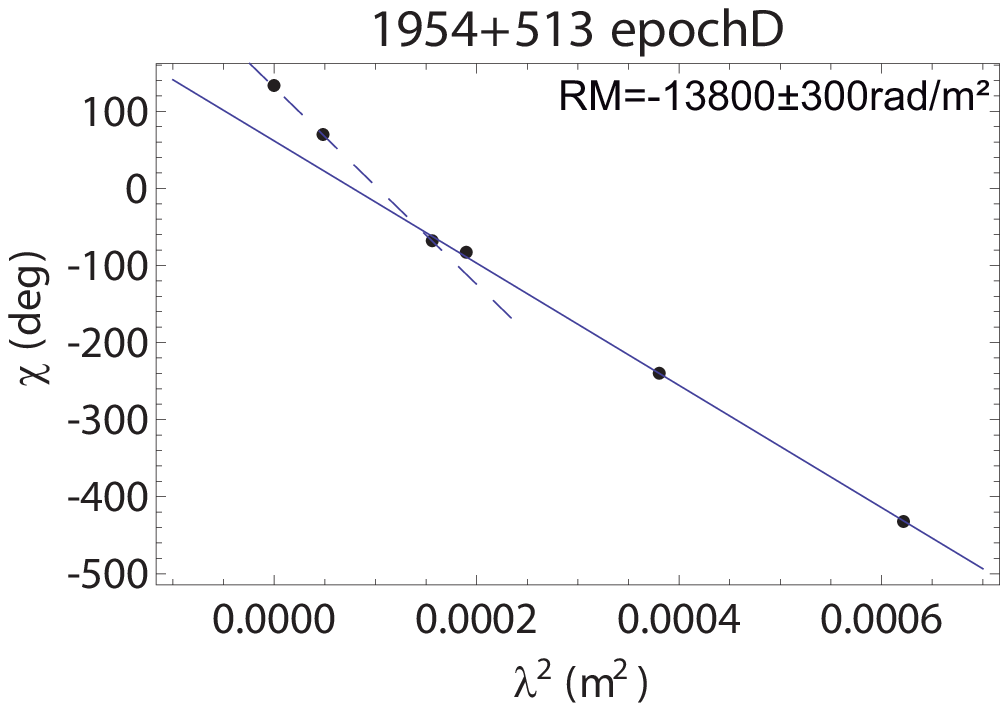}\\
\end{center}
\caption{Rotation measures for radio wavelengths and optical $\chi$'s for comparison.}
\label{fig:RM5fqs}
\end{figure*}

We essentially fixed the 43~GHz value, $\chi_{43}$, and allowed it to be rotated by no more than $\pm\pi$. Next, we used the difference between the 22 and 24~GHz $\chi$'s as a guide about both the sign and the amount of the rotation measure. That is, if $\chi_{22}$ and $\chi_{24}$ are appreciably different, this suggests the presence of relatively high Faraday rotation, which must be taken into account when fitting the lower-frequency $\chi$'s. We then apply an appropriate number of $n\pi$ rotations to $\chi_{15}$ and $\chi_{12}$ to obtain an adequate $\lambda^2$ fit. In some cases, when there was evidence that different frequencies were probing appreciably different regions, we excluded individual end points from the fits. When more than one comparably good fit was possible, we gave preference to fits that excluded fewer data points and corresponded to lower RMs.  As each of the cores is in the same optical-depth regime at all the observed frequencies, given the spectral indices (Table~4), no $\pi/2$ rotations to take into account optical-depth transitions between frequencies should be needed. 
Our results are summarized in Table~\ref{tab:5fqs-fit}. Plots of the fits given in the table are shown in Fig.~\ref{fig:RM5fqs}. The $\chi$ values for all 5 frequencies are plotted together with $\chi_{opt}$ (leftmost point in the figure) for comparison. 


Overall, it worked well to use the 22 and 24~GHz data as a guide to check for possible high rotation measures. The difference between these two $\chi$ values is equal to zero within $1\sigma$ for 0256+075, 0420$-$014, 0745+241 and 1823+568;
the two $\chi$ values differ by $1.5\sigma$ for 0133+476 and $2.5\sigma$ for 1945+513. The $\chi_{22}$ and $\chi_{24}$ values for 0906+430 and 1633+382 differ by more than $3\sigma$. Thus, the sources that are most likely to display high Faraday rotation in their cores are 0906+430, 1633+382 and 1954+513.
 
In some cases, the collected $\chi$ values for the five frequencies did not lead to a single, clearly best $\lambda^2$ fit. Some alternatives to the fits are summarized in Table~\ref{tab:altfits}, whose structure is essentially the same as Table~\ref{tab:5fqs-fit}. We briefly discuss our results for each of the eight objects below.

{\bf 0133+476.} The values for $\chi_{43}$, $\chi_{24}$ and $\chi_{22}$ agree well with a linear $\lambda^2$ dependence, indicating only a modest rotation measure in the core region. Further, it is possible to obtain a fit that is almost as good with a similar RM by including $\chi_{12}$ together with the high-frequency points; however, $\chi_{15}$ disagrees with this fit by more than $100^{\circ}$. 
We unsuccessfully tried various possibilities to see whether there was a high-rotation-measure fit that would provide an equally good fit including all 5 $\chi$ measurements. We thus propose that the 12 and 15~GHz $\chi$ values are probing a different region in the source than the higher-frequency $\chi$ values, and have used only the three highest radio frequencies for our preferred $\lambda^2$ fit; the resulting $\Delta\chi$ value is $36^{\circ}$, and the rms of the fit is less than $2^{\circ}$. 

{\bf 0256+075.} We are able to obtain an acceptable $\lambda^2$ fit for all the points except for $\chi_{43}$. This suggests we are probing a significantly different region at 43~GHz. Thus, we did not include 43~GHz in our final fit; the resulting $\Delta\chi$ value is $8^{\circ}$, and the rms of the fit is about $3^{\circ}$.

{\bf 0420--014.} There is no indication of high Faraday rotation, and we adopted the straightforward RM fit in Table~\ref{tab:5fqs-fit}, based on all 5 frequencies; the resulting $\Delta\chi$ value is $2^{\circ}$, and the rms of the fit is about $4^{\circ}$.

{\bf 0745+241.} There is no indication of high Faraday rotation, and we adopted the straightforward RM fit in Table~\ref{tab:5fqs-fit}, based on all 5 $\chi$ values; the resulting $\Delta\chi$ value is $41^{\circ}$, and the rms of the fit is about $4^{\circ}$.

{\bf 0906+430.} A good $\lambda^2$ fit is not possible using all 5 $\chi$ values, even applying reasonable $n\pi$ rotations to the lower frequencies. An acceptable fit is obtained if we exclude $\chi_{43}$, but the rms remains fairly high, about $6^{\circ}$. The implied $\Delta\chi$ value in this case is $2^{\circ}$. The three highest-frequency $\chi$ values are very well aligned, suggesting that these higher frequencies probe a different region than the lower two frequencies, and so should be fit on their own. In this way, the resulting $\Delta\chi$ value is $86^{\circ}$, and the rms of the fit is about $2^{\circ}$. Overall, it is difficult to be sure which of these should be preferred, and both options are given in Table~\ref{tab:5fqs-fit} and shown in Fig.~\ref{fig:RM5fqs}.

{\bf 1633+382.} Inspection of Table~\ref{tab:5fqs-fit} makes it immediately clear that it is not possible to obtain an acceptable $\lambda^2$ fit with the nominal measured $\chi$ values. Further, the large difference between $\chi_{22}$ and $\chi_{24}$ suggests a fairly large core RM. Applying rotations of $+\pi$ to the $\chi_{15}$ and $+3\pi$ to $\chi_{12}$ yields a good fit to all the frequencies except for 15~GHz; at the same time, the parameters for the fits including and excluding $\chi_{15}$ are fairly similar. In addition, these fits are close to those for the fit that is obtained using only the $\chi$ values at the three highest frequencies. We adopt the three-frequency fit as our preferred fit, assuming that the increased rms obtained when all five frequencies are used reflects the fact that the lower-frequency data are probing somewhat different regions within the source. The resulting $\Delta\chi$ value is $33^{\circ}$, and the rms of the fit is about $3^{\circ}$.
 
{\bf 1823+568.} There appears to be a clear misalignment between $\chi_{43}$ and the rest of the frequencies. An excellent $\lambda^2$ fit is obtained using the remaining four $\chi$ values. This suggests that we are beginning to sample a significantly different region at 43~GHz. Thus, we did not include 43~GHz measurement when fitting the $\chi$'s for this core. The resulting $\Delta\chi$ value is $60^{\circ}$, and the rms on the fit is less than $1^{\circ}$.

{\bf 1954+513.} $\chi_{43}$ lies substantially above the $\lambda^2$ fit obtained using all five $\chi$ values. If we obtain a fit excluding this point, the resulting RM differs only slightly, but the RMS decreases by about a factor of four. We accordingly adopt the fit without $\chi_{43}$. The resulting $\Delta\chi$ value is $80^{\circ}$, and the rms of the fit is about $3^{\circ}$. It is intriguing that $\chi_{opt}$, $\chi_{43}$ and $\chi_{24}$ are well aligned (dashed in Fig. 1), but this may be a coincidence, since $\chi_{22}$ is offset somewhat from this line. If it is not a coincidence, then the excellent alignment of $\chi_{opt}$, $\chi_{43}$ and $\chi_{24}$ obviously implies $\Delta\chi \approx 0$, with $RM=-22300\pm600$ rad/m$^2$.

\begin{landscape}
\begin{center}
\begin{table}
\caption{Measured $\chi$ values and results for $\chi$ vs. $\lambda^2$ fits}
\centering
\begin{tabular}{ c c c c c c c c c c c }
\hline\hline
Source& $\chi_{12GHz}(^\circ)$& $\chi_{15GHz}(^\circ)$& $\chi_{22GHz}(^\circ)$& $\chi_{24GHz}(^\circ)$&$\chi_{43GHz}(^\circ)$ & $\chi_{0}(^\circ)$ & RM (rad/m$^2$)&RMS& $\chi_{opt}(^\circ)$ & $\Delta\chi(^\circ)$ \\
$[1]$ & [2] & [3] & [4]& [5] & [6] & [7]& [8] & [9] & [10]& [11]\\
\hline
0133+476$^{\ddag}$ & $-43\pm4$ & $-67\pm4$ & $81\pm4$ & $90\pm4	$  & $103\pm4$  & $111\pm4$  & $-2500\pm500$  & $1.5$ & $75\pm1$ &36\\
0256+075$^{\dag}$  & $48\pm4$  & $19\pm4$ & $8\pm4$ & $	7\pm4	$  & $33\pm5$   & $-9\pm5$   & $1530\pm200$   & $2.7$ & $-17\pm1$ &8\\
0420--014	   & $64\pm4$  & $42\pm4$ & $35\pm4$ & $41\pm4	$  & $34\pm5$   & $29\pm5$   & $880\pm220$    & $4.1$ & $27\pm1$ &2\\
0745+241	   & $133\pm4$ & $103\pm4$ & $61\pm4$ & $63\pm4	$  & $56\pm4$   & $42\pm5$   & $2550\pm270$   & $4.4$ & $83\pm1$ &41\\
0906+430$^{\dag}$  & $9\pm5$   & $65\pm7$ & $87\pm5$ & $105\pm4	$  & $184\pm4$  & $129\pm9 $ & $-3200\pm500$  & $5.9$ & $131\pm3$ &2\\
0906+430$^{\ddag}$ & $9\pm5$   & $65\pm7$ & $87\pm5$ & $105\pm4	$  & $184\pm4$  & $37\pm6$   & $-12100\pm700$ & $2.4$ & $131\pm3$ &86\\
1633+382$^{\ddag}$ & $78\pm4$  & $27\pm4$ & $77\pm4$ & $42\pm6	$  & $-100\pm21$  & $20\pm8$  & $22040\pm960$  & $3.3$ & $-13\pm1$ &33\\
1823+568$^{\dag}$  & $23\pm4$  & $5\pm4	$ & $-9\pm4$ & $-10\pm4	$  & $0\pm4$    & $-22\pm1$  & $1250\pm30$    & $0.3$ & $38\pm1$ &60\\
1954+513$^{\dag}$  & $-72\pm4$ & $-60\pm4$ & $-83\pm4$ & $-68\pm4$ & $-110\pm4$ & $86\pm16$  & $-13800\pm300$ & $3.2$ & $146\pm5$ &80\\
\hline
\multicolumn{11}{l}{Columns show [1] the source name, [2]--[6] nominal $\chi$ values in degrees, [7] $\chi_0$ in degrees, [8] RM in rad/m$^2$, [9] rms of the $\lambda^2$ fit, }\\
\multicolumn{11}{l}{[10] optical polarization angle and [11] $\Delta\chi=|\chi_{opt}-\chi_0|$. }\\
\multicolumn{11}{l}{$^{\dag}$ \footnotesize{$\chi_{43GHz}$ was not used in the fit. See text for details.}}\\
\multicolumn{11}{l}{$^{\ddag}$ \footnotesize{$\chi_{12GHz}$ and $\chi_{15GHz}$ were not used in the fit. See text for details.}}\\
\end{tabular}\\
\begin{center}
\end{center}  
\label{tab:5fqs-fit}
\end{table}


\begin{table}
\caption{Alternative possibilities for RM fits}
\centering
\begin{tabular}{ c c c c c c c c c c c }
\hline\hline
Source & Case & $\chi_{12GHz}(^\circ)$& $\chi_{15GHz}(^\circ)$& $\chi_{22GHz}(^\circ)$& $\chi_{24GHz}(^\circ)$&$\chi_{43GHz}(^\circ)$ & $\chi_{0}(^\circ)$ & RM (rad/m$^2$)& RMS& $\Delta\chi(^\circ)$ \\
$[1]$ & [2] & [3] & [4]& [5] & [6] & [7]& [8] & [9] & [10]& [11]\\
\hline

0133+476& A 	&$	-43\pm4		$&$	-67\pm4	$&$	81\pm4	$&$	90\pm4	$&$	103\pm4	$&$	120\pm35	$&$	-5400\pm1800	$&$	 27.9 	$&	 45\\	
	& B 	&$	-223\pm4	$&$	-67\pm4	$&$	81\pm4	$&$	90\pm4	$&$	103\pm4	$&$	168\pm22	$&$	-10700\pm1100	$&$	 18.7 	$&	 93\\	
	& C 	&$	677\pm4		$&$	473\pm4	$&$	261\pm4	$&$	270\pm4	$&$	103\pm4	$&$	84\pm22 	$&$	17100\pm1100	$&$	 20.4 	$&	 9\\	
	& D 	&$	-223\pm4	$&$	-67\pm4	$&$	81\pm4	$&$	90\pm4	$&$	-	$&$	23\pm9  	$&$	-12000\pm400	$&$	6.0	$&	 128\\	
	& E 	&$	-43\pm4		$&$	 -         	$&$	81\pm4	$&$	90\pm4	$&$	103\pm4	$&$	126\pm7 	$&$	-4700\pm400 	$&$	6.1	$&	 51\\	
	& F$^{\dag}$ 	&$	 - 	$&$	-	$&$	81\pm4	$&$	90\pm4	$&$	103\pm4	$&$	111\pm4 	$&$	-2500\pm500	$&$	1.5	$&	 36\\	
\hline
0906+430& A	&$	9\pm5		$&$	65\pm7	$&$	87\pm5	$&$	105\pm4	$&$	184\pm4	$&$	167\pm22	$&$	-5000\pm1400	$&$	19.8	$&	36\\
	& B	&$	-351\pm5	$&$	-115\pm7	$&$	87\pm5	$&$	105\pm4	$&$	184\pm4	$&$	67\pm13	$&$	-16500\pm800	$&$	11.4	$&	64\\
	& C$^{*}$	&$	9\pm5	$&$	65\pm7	$&$	87\pm5	$&$	105\pm4	$&$	-	$&$	129\pm9	$&$	-3200\pm500	$&$	5.9	$&	2\\
	& D$^{*}$	&$	-	$&$	-	$&$	87\pm5	$&$	105\pm4	$&$	184\pm4	$&$	37\pm6	$&$	-12100\pm700	$&$	2.4	$&	86\\
\hline
1633+382& A	&$	78\pm4		$&$	27\pm4	$&$	77\pm4	$&$	42\pm5	$&$ -100\pm21	$&$	26\pm27	$&$	22460\pm1350	$&$	27	$&	39\\
	& B	&$	78\pm4		$&$	27\pm4	$&$	77\pm4	$&$	42\pm5	$&$ -           $&$    -48\pm14 $&$	28700\pm650	$&$	9.7	$&35\\
	& C$^{\dag}$	&$	-	$&$	-	$&$	77\pm4	$&$	42\pm5	$&$    -100\pm21 $&$	20\pm8	$&$	22040\pm960	$&$	3.3	$&33\\
\hline
1954+513& A	&$	-72\pm4		$&$	-60\pm4	$&$	-83\pm4	$&$	-68\pm4	$&$	-110\pm4$&$	86\pm16	$&$	-14800\pm800	$&$	13.3	$&	60\\
	& B$^{\dag}$&$	-72\pm4		$&$	-60\pm4	$&$	-83\pm4	$&$	-68\pm4	$&$	-	$&$	66\pm6	$&$	-13800\pm300	$&$	3.2	$&	80\\
	& C	&$	-792\pm4	$&$	-420\pm4$&$	-83\pm4	$&$	-68\pm4	$&$	70\pm4	$&$	174\pm20$&$	-27000\pm1000	$&$	15.6	$&	25\\
\hline
\multicolumn{11}{l}{$^{\dag}$ \footnotesize{Preferred fit.}}\\
\multicolumn{11}{l}{$^*$ \footnotesize{Both fits comparable.}}
\end{tabular}\\
\centering
\label{tab:altfits}
\end{table}
\end{center}
\end{landscape}


\section{Discussion}

\subsection{The Optical and VLBI Core $\chi$ Values}

The five-frequency data have partially resolved the issues we had with previous three-frequency data: i) we can essentially rule out internal Faraday rotation as the cause of any misalignments between $\chi_{opt}$ and $\chi_0$ and ii) we now have a better basis to determine cases of high Faraday rotation in the VLBA cores. If any of the cores were undergoing internal Faraday rotation, both an oscillation of $\chi$ in the observed frequency range and a characteristic frequency dependence of the degree of polarization would be expected, but are not observed (see Paper~I). Comparison of $\chi_{22}$ and $\chi_{24}$, has proved to be a good (although not perfect) tool to identify possible high Faraday rotation.

Although we are able to obtain reasonable $\lambda^2$ fits using all 5 frequencies in some cases (0420$-$014, 0745+241), $\chi_{43}$ deviates from the linear behaviour shown by the other 
frequencies in several other sources (0256+075, 1823+568, 1954+513 and possibly 0906+430). This can be understood if the 43-GHz VLBA observations are probing a significantly different region than
the lower-frequency observations. Similarly, in some sources, $\chi_{43}, \chi_{24}$ and $\chi_{22}$ are well aligned, but $\chi_{15}$ and $\chi_{12}$ are offset from this linear relation
(0133+476, possibly 0906+430). 

Upon inspection of Table~\ref{tab:5fqs-fit}, it becomes clear that there is no obvious pattern for $\Delta\chi$: two sources have $\Delta\chi<10^{\circ}$ (0256+075, 0420$-$014), one has
$\Delta\chi \geq 80^{\circ}$ (1954+513), and the $\Delta\chi$ value for 0906+430 is most likely close to $90^{\circ}$, but may be close to $0^{\circ}$. The remaining four $\Delta\chi$ values
are all between $30^{\circ}$ and $60^{\circ}$. There is no obvious pattern for $\Delta\chiö'=|\chi_{opt}-\chi_{43}|$ when $\chi_{43}$ is not aligned with other radio frequencies. Thus, the additional information provided by the increased frequency coverage of our new data set has not led to the emergence of a clear peak in the $\Delta\chi$ distribution for the 8 AGN considered here.
 
This has a number of implications. First, this supports the preliminary conclusion of Paper~I that not all sources display aligned optical and VLBA core polarization angles. Second, there may be some factor differentiating the sources in our total sample with $\Delta\chi \approx 0$ from those for which $\Delta\chi$ is far from zero; e.g., they may be intrinsically different, or there may be some physical or observational process leading to this difference. Third, the $\Delta\chi$ behaviour shown by a particular object could in principle change with time. For example, D'Arcangelo et. al (2007) showed that the optical and radio core position angles of 0420--014 followed each other fairly well over several days, but we have no information about whether this continues over long periods of months or years. In the following sections, we will consider these issues.

\subsection{Time Variation of $\Delta\chi$}

The observations discussed in Paper I and those presented here yield a subset of sources for which we have $\Delta\chi$ measurements for two epochs; these results are summarized in Table~7.

\begin{table*}
\begin{center}
\caption{Time evolution of core RM and $\Delta\chi$}
\begin{tabular}{c c c c c c c c }
\hline\hline
Source & Epoch 1 & Epoch 2 & $\Delta t$ (yr) & RM (Ep. 1) (rad/m$^2$)& RM (Ep. 2) (rad/m$^2$)& $\Delta\chi$(Ep. 1) ($^\circ$)& $\Delta\chi$(Ep. 2) ($^\circ$)\\
$[1]$&[2]&[3]&[4]&[5]&[6]&[7]&[8]\\
\hline
0133+476&	C	&	D	&	3.10	&$	200\pm30	$&$	-2500\pm500	$&	59	&	36	\\
0256+075&	A	&	D	&	4.01	&$	200\pm820	$&$	1530\pm200	$&	60	&	8	\\
0420--014&	B	&	D	&	3.64	&$	270\pm200	$&$	880\pm220	$&	43	&	2	\\
0745+241&	B	&	D	&	3.64	&$	-1780\pm320	$&$	2550\pm270	$&	51	&	41	\\
0906+430$^{\dagger}$&A	&	D	&	4.01	&$	-24440\pm470	$&$	-3200\pm500	$&	8	&	2	\\
0906+430$^{\ddagger}$&A	&	D	&	4.01	&$	-24440\pm470	$&$	-12100\pm700	$&	8	&	86	\\
1633+382&	A	&	D	&	4.01	&$	-570\pm430	$&$	22040\pm960	$&	29	&	40	\\
1823+568&	GA	&	D	&	6.24	&$	400\pm100	$&$	1250\pm30	$&	54	&	60	\\
1954+513&	B	&	D	&	3.64	&$	-24400\pm640	$&$	-13800\pm300	$&	4	&	60	\\
OJ287&	GB	&	C	&	2.56	&$	-	$&$	-1920\pm120	$&	3	&	68	\\
3C279&	GB	&	B	&	2.03	&$	-2360\pm120	$&$	-1020\pm110	$&	3	&	48	\\
\hline
\multicolumn{8}{l}{Columns are as follows: [1] source name, [2]--[3] epochs observed (GA: August 7, 2004; GB: March 5, 2003; A: November 1, 2004;}\\
\multicolumn{8}{l}{B: March 15, 2005; C: September 26, 2005; and D: November 2, 2008), [4] time (in years) between those two epochs, [5]--[6] derived} \\
\multicolumn{8}{l}{RMs and [7]--[8] the two $\Delta\chi$ values at the corresponding epochs.}\\
\multicolumn{8}{l}{$^\dagger$ Fit obtained using the $\chi$ values at 12, 15, 22 and 24 GHz.}\\
\multicolumn{8}{l}{$^\ddagger$ Fit obtained  using the $\chi$ values at 22, 24 and 43 GHz.}
\end{tabular}
\end{center}
\label{tab:twoepochs}
\end{table*}

The $\Delta\chi$ values inferred for the two epochs are roughly the same for 0133+576, 0745+241, 1633+382, 1823+568 and possibly 0906+430, whereas the $\Delta\chi$ values for the other sources differ by more than about $20^{\circ}$. Thus, roughly half of these sources display a substantial change in their $\Delta\chi$ estimates over time scales of several years. Comparing the core RM fits for the two epochs (columns [5] and [6]), we do not see any clear trend for systematically higher or lower RM values in our five-frequency data (epoch D), compared to the three--frequency data (epochs GA, GB, A, B, C). Overall, the RM values are similar to order of magnitude, with the exception of 1633+382, which differ by a factor of 50. 
This suggests that there were no systematic problems with our previous three-frequency RM fits, and that those fits were generally reliable despite the limited frequency coverage.

The signs of the core rotation measures for 0133+476, 0745+241 and 1633+382 apparently changed between epochs. However, this is not unprecedented. Zavala \& Taylor (2001) found similar changes in the core rotation measures in 3C273 and 3C279. 

If $\Delta\chi$ changes over time, we can consider several possibilities. First, it might be that $\chi_{opt}$ and $\chi_0$ each change independently, such that $\Delta\chi$ varies randomly. It could also be that $\Delta\chi$ follows some predictable path, such as an oscillation around a certain value or a rotation over time. Monitoring of $\chi_{opt}$ and $\chi_{VLBA}$ over a longer time period would be required to test this. A third option would be that $\Delta\chi$ is correlated with some other physical properties of the source -- for example, that $\Delta\chi$ is close to zero only when the core is active, or quiescent. We consider this last possibility in the following sections.

\subsection{Other Physical Properties and $\Delta\chi$}

In principle, a characteristic difference between the optical and the Faraday-corrected VLBA core polarization angle could arise as the result of some physical properties of the core region. We considered several possibilities in Paper I. For example, we found no evidence for a correlation between $\Delta\chi$ and $|\chi_0-\Theta_j|$, where $\Theta_j$ is the inner jet direction. 
No evidence was found for a correlation between the degree of polarization in the core and $\Delta\chi$, although there were slight indications that $\Delta\chi$ varies over a smaller range of values as the degree of optical polarization $m_{opt}$ increased. Finally, we found no link between depolarization and $\Delta\chi$.

We consider two further possible relations between $\Delta\chi$ and other physical characteristics of the source, namely, the intrinsic core RM and the core-region magnetic field. For completeness, we will do this for all available epochs, including those analyzed in Paper I.

\subsubsection{Rotation Measure}

\begin{figure}
\begin{center}
\includegraphics[width=8cm]{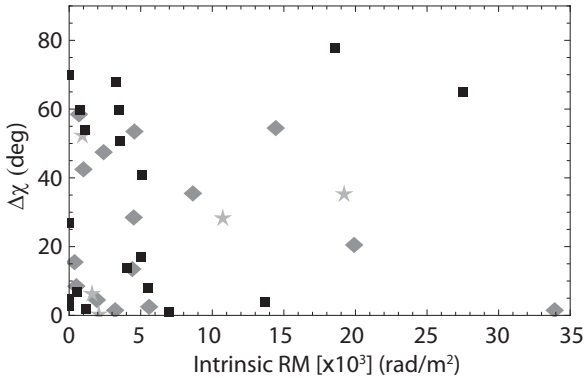}
\end{center}
\caption{Intrinsic rotation measure RM compared against $\Delta\chi$. Black squares, dark gray diamonds and light gray stars indicate BL Lac objects, HPQs and LPQs respectively. Note that, for clarity, 0906+430, 2230+014 for epoch A and 1954+413 for epoch D have not been plotted as their intrinsic RM is higher than 60 000 rad/m$^2$.}
\label{fig:DC_VS_RM}
\end{figure}

We first look for a correlation between the rotation measures found for our cores and $\Delta\chi$. For sources with known redshift, intrinsic rest frame core rotation measures can be calculated using $RM_{int}=RM_{obs}(1+z)^2$. 
The results are given in Fig.~\ref{fig:DC_VS_RM}, where black squares indicate BL Lac objects; dark gray diamonds high polarized quasars (HPQs) and light grey stars low polarized quasars (LPQs) (for a complete list of the RMs, see Paper I). It is clear that no correlation appears. The magnitude of the rotation measure has no obvious relationship to the observed $\Delta\chi$.

\subsubsection{Core-region Magnetic Field}

Typically, due to the finite resolution of the observations, the VLBI ``core'' is in fact a combination of the true optically thick core (i.e., the surface with optical depth $\tau=1$) plus the unresolved innermost region of the optically thin jet. Due to the frequency dependence of the $\tau=1$ surface (K\"onigl 1981), the position of the radio core varies as $r\propto\nu^{-1/k_r}$, with $r$ the distance from the central engine and $k_r=((3-2\alpha)m+2-2)/(5-2\alpha)$, where $m$ and $n$ describe the decay of the magnetic field ($B=B_{1}r^{−m}$) and particle number density ($N=N_{1}r^{−n}$) with distance from the jet base (i.e., the apex of the cone formed by the jet). In the case of equipartition, $k_r=1$, independent of the spectral index $\alpha$, with the choices of $m=1$, $n=2$ being physically reasonable.

We can derive the magnetic field strength from this core shift combining the results of Lobanov (1998) and Hirotani (2005) (also see O'Sullivan \& Gabuzda 2009). The magnetic field in Gauss at 1~pc from the jet base is given by
\begin{equation}
B_1\sim0.025 \left(\frac{\Omega_{r\nu}^3 (1+z)^2}{\delta\phi\sin^2\theta}\right)^{1/4},
\end{equation}
where $z$ is the redshift, $\delta$ the Doppler factor, $\phi$ the half opening angle of the jet, $\theta$ the viewing angle and $\Omega_{r\nu}$ the core position offset, defined by Lobanov (1998) as
\begin{equation}
\Omega_{r\nu}=4.85\times10^{-9}\frac{\Delta r_{mas}D_L}{1+z^2}\left(\frac{\nu^{1/k_r}_1 \nu^{1/k_r}_2}{\nu^{1/k_r}_2-\nu^{1/k_r}_1}\right),
\end{equation}
with $D_L$ being the luminosity distance and $\Delta r_{mas}$ the core shift in mas found between the frequencies $\nu_1$ and $\nu_2$.

\begin{figure*}
\begin{center}
\includegraphics[width=6.8cm]{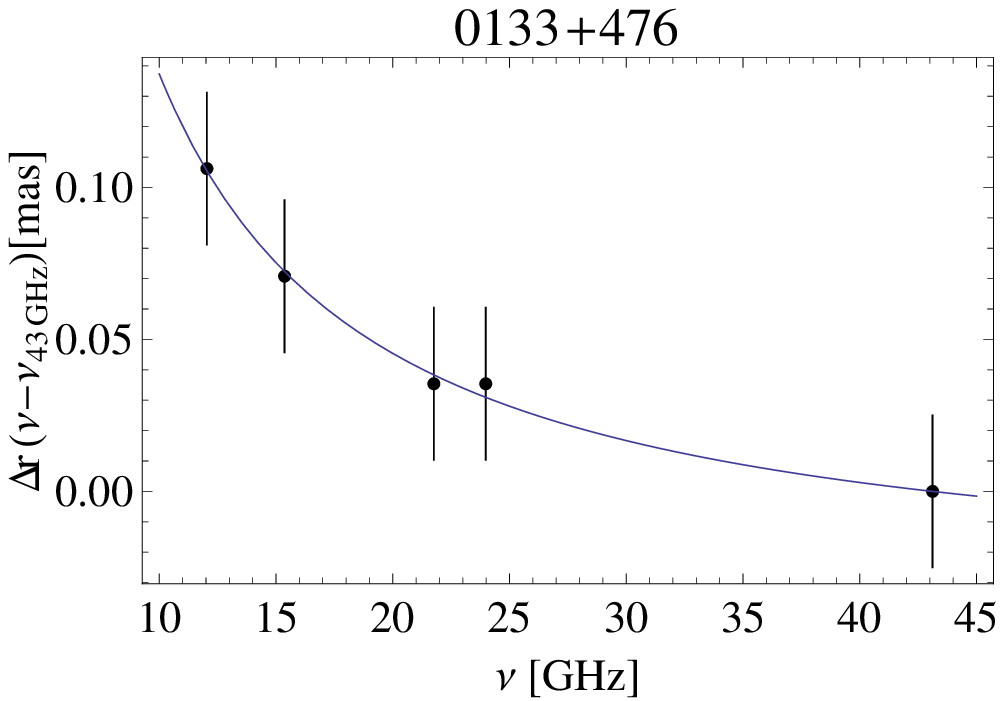}\hspace{0.8cm}
\includegraphics[width=6.8cm]{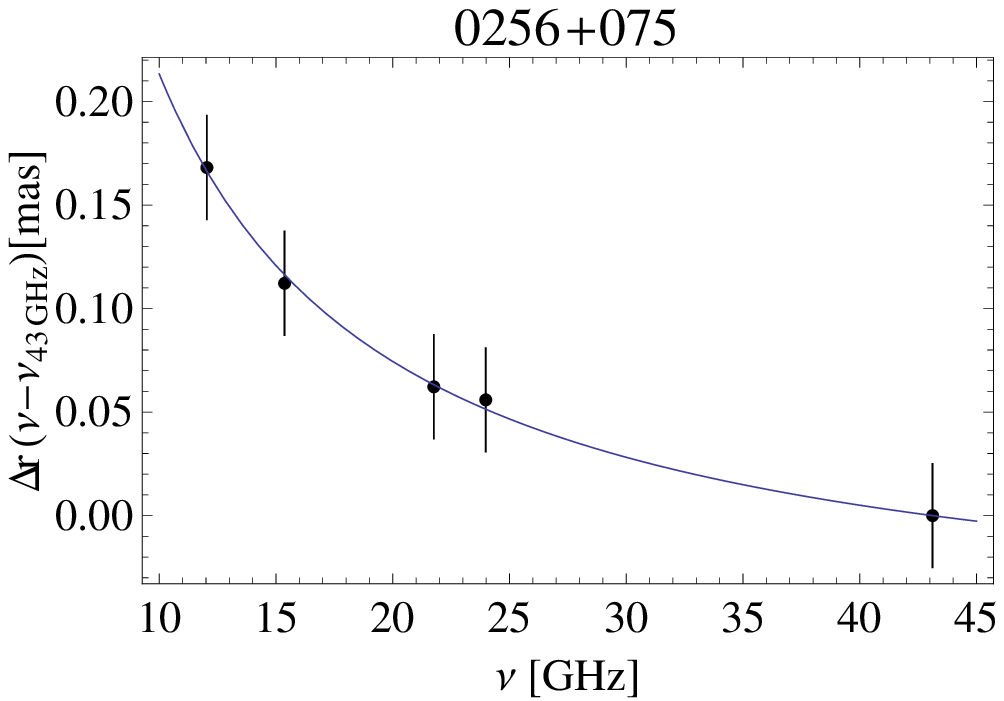}\\
\includegraphics[width=6.8cm]{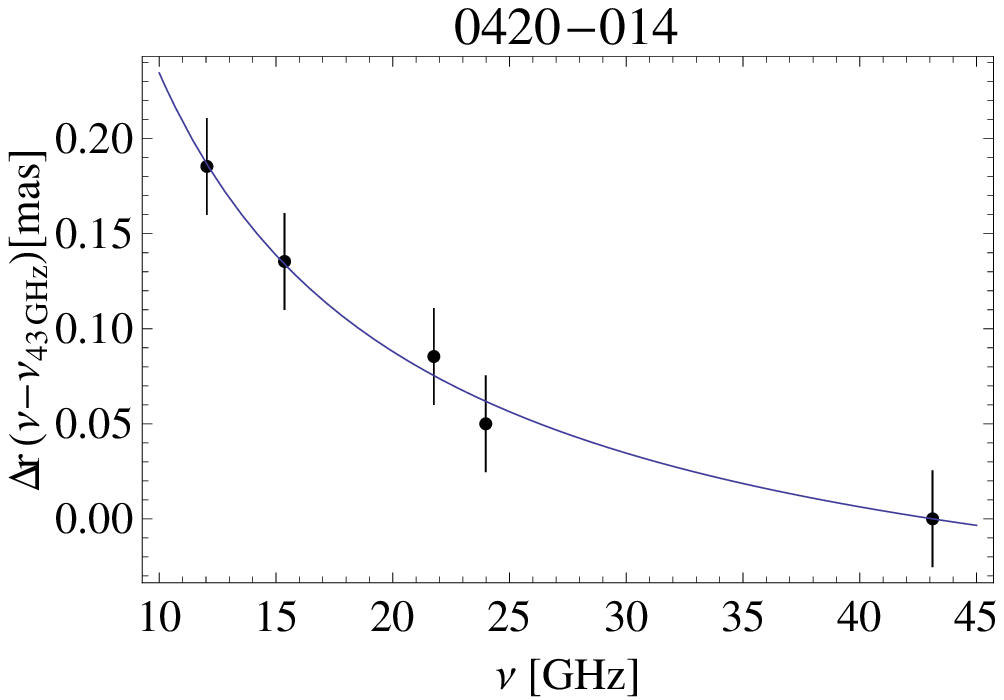}\hspace{0.8cm}
\includegraphics[width=6.8cm]{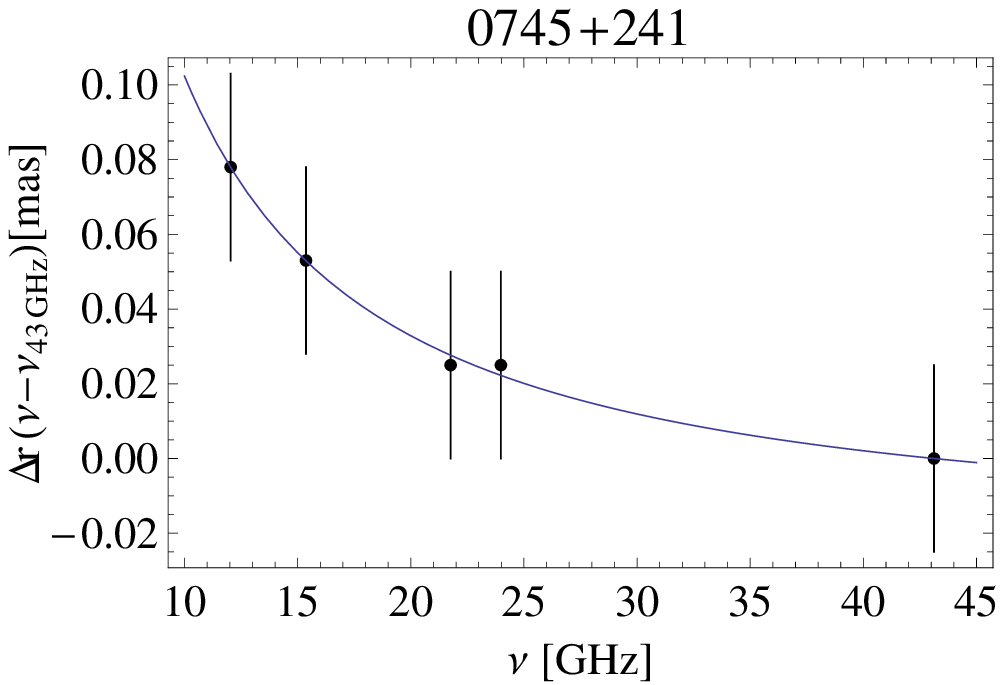}\\

\includegraphics[width=6.8cm]{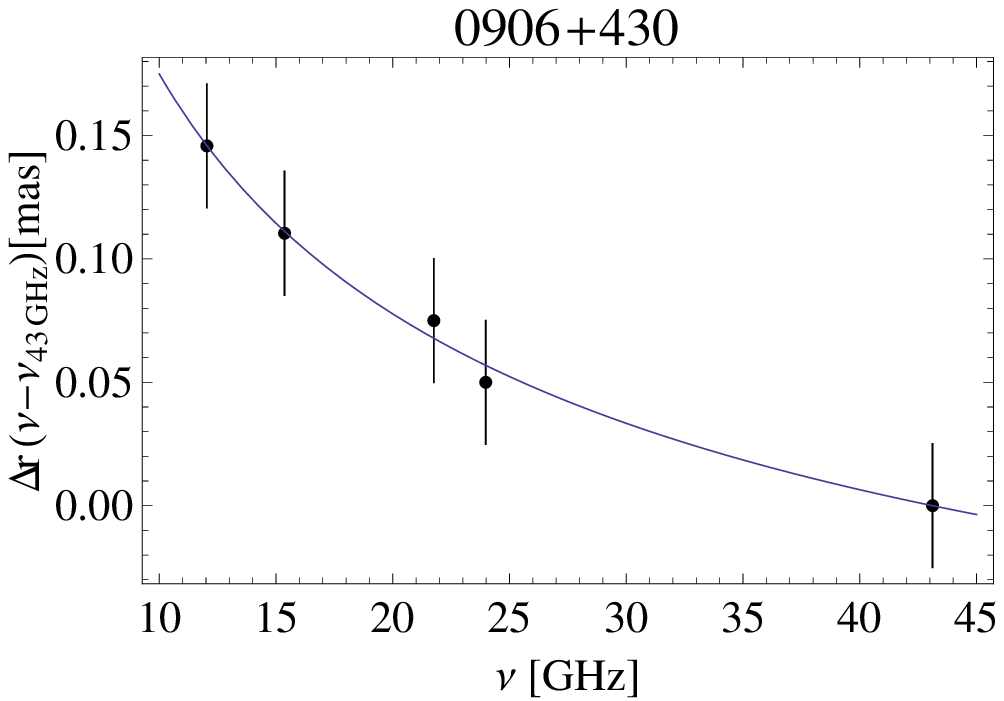}\hspace{0.8cm}
\includegraphics[width=6.8cm]{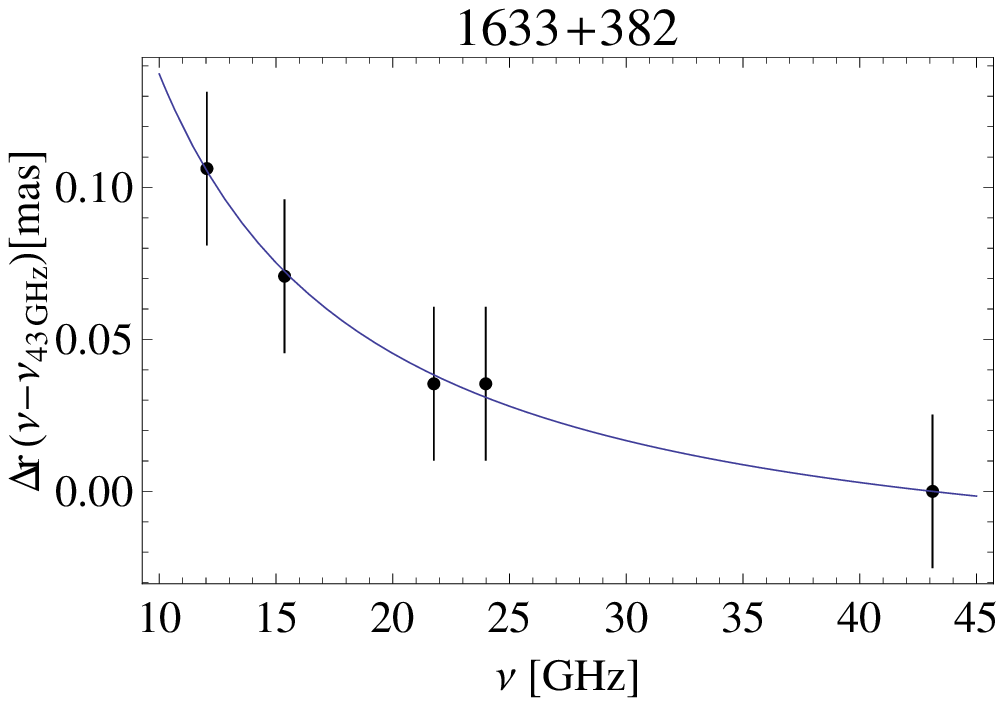}\\
\includegraphics[width=6.8cm]{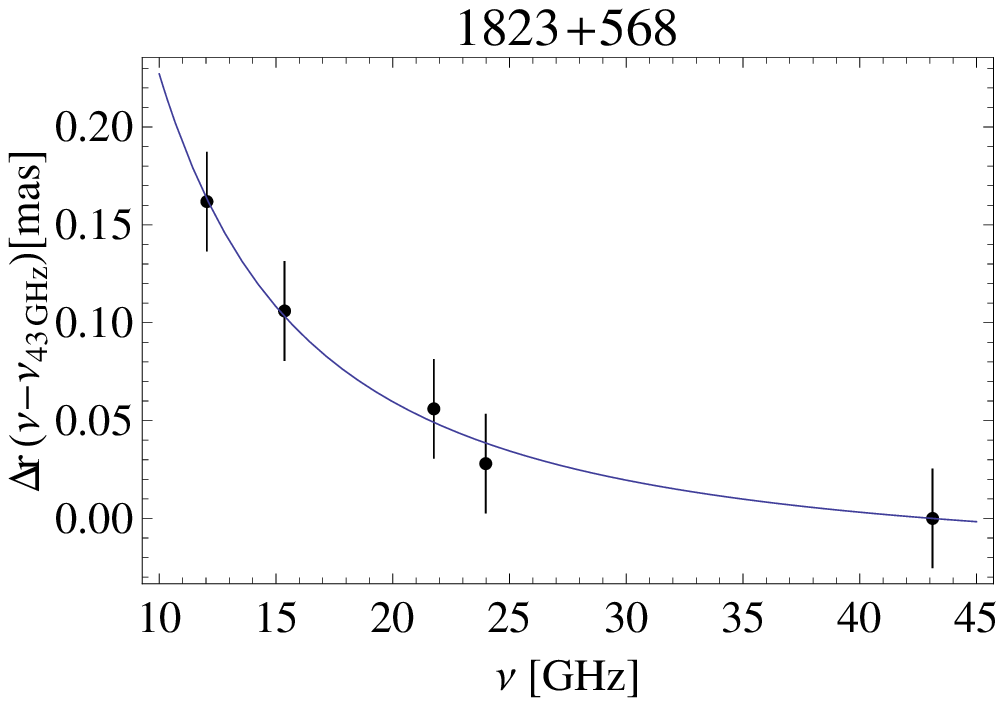}\hspace{0.8cm}
\includegraphics[width=6.8cm]{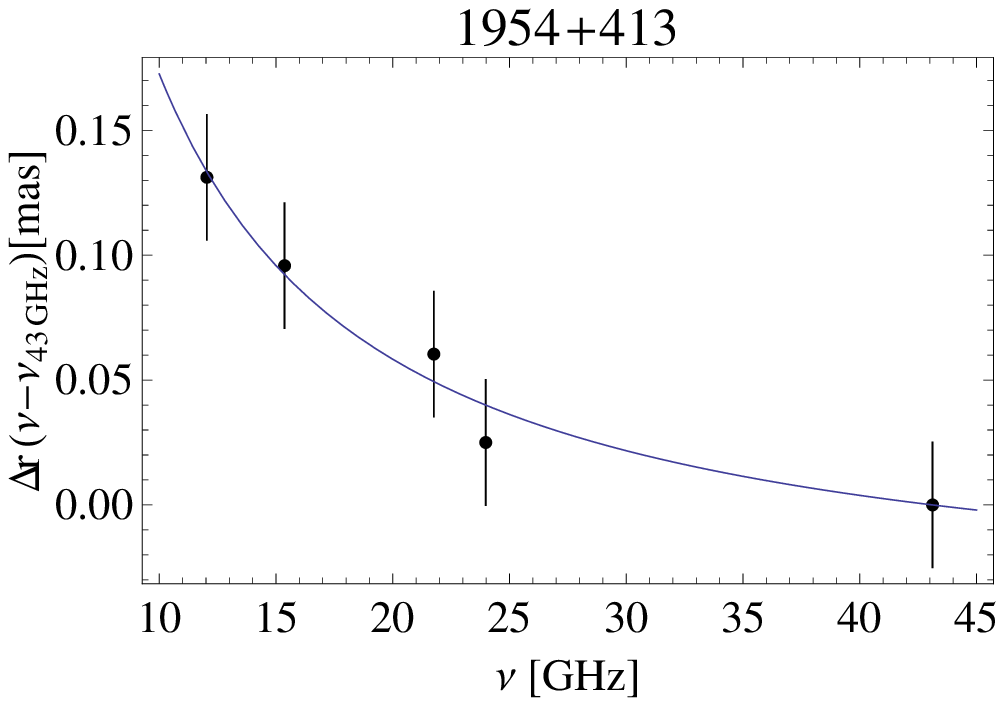}\\
\end{center}
\caption{Frequency dependent shift of the core position for the sources studied. Fits correspond to the spectral shift power index $k_r$ when 43 GHz is used as the reference frequency.}
\label{fig:kr}
\end{figure*}

We have estimated $k_r$ for the sample using an exponential of the form $\Delta r=A\times(\nu^{-1/k_r}-(43)^{-1/k_r})$, where $\Delta r$ are the core shifts given in Table~3 and 43~GHz is used as the reference frequency. Fits are shown in Figure~\ref{fig:kr} and the corresponding $k_r$ values are given in Table~\ref{tab:magnetic_stuff}.  For all but two sources (0906+430 and 1823+568), $k_r$ is equal to 1 within the $1\sigma$ uncertainties, and for only one (1823+568) is $k_r\neq1$ within $2\sigma$. Thus, the values obtained for $k_r$ are mostly compatible with equipartition in the core region.

Once the core shifts and $k_r$ have been obtained, the core position offset $\Omega_{r\nu}$ can be calculated. As the majority of the cores are found to be close to equipartition, the magnetic field $B$ at 1~pc from the jet base can be derived, provided that there are adequate values for the Doppler factor, the jet opening angle, the viewing angle and the redshift.  
These parameters and the inferred results are also given in Table~\ref{tab:magnetic_stuff}. 
In all cases, a jet opening angle $\phi=0.5^{\circ}$ is assumed.

\begin{table*}
\caption{$k_r$, $\Omega_{r\nu}$, and $B_1$}
\begin{center}
\begin{tabular}{ c c c c c c c c c c }
\hline\hline
Source & $k_r$ & $z$ & $D_L$ (Mpc) & $\phi(^{\circ})$ & $\Gamma^{\dagger}$ & $\theta(^{\circ})^{\dagger}$ & $\delta^{\dagger}$ & $\Omega_{r\nu}$ (pc/Hz) & $B_{1}$ (G)\\ 
$[1]$&[2]&[3]&[4]&[5]&[6]&[7]&[8]&[9]&[10] \\
\hline 
0133+476&$0.9\pm0.1$&0.86&$5.49\times10^{3}$&0.5&14.4&2.5&20.7&$(1.47\pm0.04)\times10^{10}$&$0.3\pm0.3$\\
0256+075&$1.0\pm0.1$&0.89&$5.60\times10^{3}$&0.5&-&-&-&$(2.20\pm0.20)\times10^{10}$&$-$\\
0420-014&$1.2\pm0.4$&0.92&$5.93\times10^{3}$&0.5&11.4&1.9&19.9&$(2.11\pm0.06)\times10^{10}$&$0.8\pm0.3$\\
0745+241&$0.9\pm0.1$&0.41&$2.22\times10^{3}$&0.5&-&-&-&$(0.76\pm0.03)\times10^{10}$&$-$\\
0906+430&$2.2\pm1.0$&0.67&$6.83\times10^{3}$&0.5&-&-&-&$(2.60\pm0.60)\times10^{10}$&$-$\\
1633+382&$0.9\pm0.1$&1.81&$1.39\times10^{4}$&0.5&30.5&2.5&21.5&$(1.63\pm0.04)\times10^{10}$&$0.7\pm0.3$\\
1823+568&$0.6\pm0.1$&0.66&$3.99\times10^{3}$&0.5&37.8&5.0&6.4&$(1.70\pm0.60)\times10^{10}$&$0.7\pm0.2$\\
1954+413&$0.9\pm0.4$&1.22&$8.53\times10^{3}$&0.5&-&-&7.4&$(1.40\pm0.30)\times10^{10}$&$-$\\
\hline 
\multicolumn{10}{l}{Columns are as follows: source name, spectral shift power index $k_r$; redshift $z$; luminosity distance $D_L$ in Mpc;}\\\multicolumn{10}{l}{Lorentz factor $\Gamma$, viewing angle $\theta$, Doppler factor $\delta$, core shift parameter $\Omega_{r\nu}$ and the inferred magnetic field at } \\
\multicolumn{10}{l}{1~pc from the jet base $B_{1}$. }\\
\multicolumn{10}{l}{$^\dagger$ from Hovatta  et al (2009)}
\end{tabular}\\
\end{center}
\label{tab:magnetic_stuff}
\end{table*}

Values of $B_{1}$ were obtained for four sources 
(Table~\ref{tab:magnetic_stuff}). These values are typically $\simeq 0.7$~Gauss, similar to previous results for other radio-loud AGN (Lobanov 1998, O'Sullivan \& Gabuzda 2009).  
There is no tendency for cores displaying $k_r$ values far from unity (indicating that they are relatively far from equipartition) to stand out in terms of their $\Delta\chi$ values (Figure 4).

\begin{figure}
\begin{center}
\includegraphics[width=8cm]{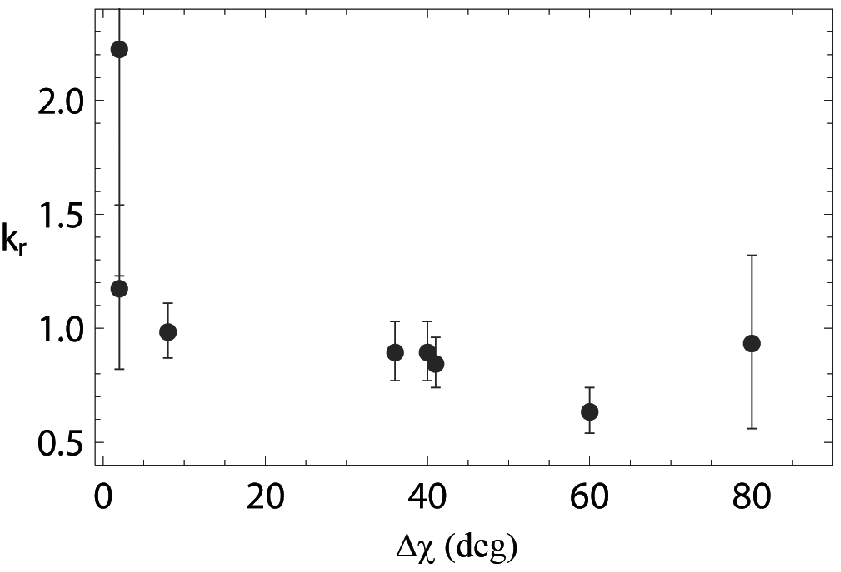}
\end{center}
\caption{Spectral shift power index $k_r$ parameter for the $\Delta\chi$ values obtained in the sources studied. No tendency in $\Delta\chi$ is seen for these cores with $k_r$ differing from unity.}
\label{fig:DC_vs_Kr}
\end{figure}

A check was made for any dependence of $\Delta\chi$ on $\Omega_{r\nu}$, as derived from the core shifts, and on the strength of the magnetic fields deduced. However, there are five-frequency data for only eight sources, and we have sufficient information to derive magnetic fields for only half of these; too few sources to be of any statistical value. This issue can be partially addressed by deriving $\Omega_{r\nu}$ (and hence, magnetic fields) for the sources studied in Paper I. However, those data were obtained at only three frequencies (15, 22 and 43~GHz), which proved to be insufficient for reliable fits for $k_r$. Therefore, based on our five-frequency results and the results of O'Sullivan \& Gabuzda (2009), we assume equipartition ($k_r=1$) for the observations in Paper I; this assumption is likely to be incorrect only for a small minority of those sources. The equivalent of Table~\ref{tab:magnetic_stuff} for the data of Paper I is summarized in Table \ref{tab:Bfield3freqs}, where sources have been marked as belonging to epochs A (November 1, 2004), B (March 15, 2005) or C (September 26, 2005).

\begin{table*}
\caption{Magnetic-Field Parameters for Sources in Paper I (assuming $k_r=1$)}
\begin{center}
\begin{tabular}{ c c c c c c c c c }
\hline\hline
Source & Epoch & $z$ & $D_L$ (Mpc) & $\Gamma^{\dagger}$ & $\theta(^{\circ})^{\dagger}$ & $\delta^{\dagger}$ & $\Omega_{r\nu}$ (pc/Hz) & $B_{1}$ (G)\\ 
$[1]$&[2]&[3]&[4]&[5]&[6]&[7]&[8]&[9] \\
\hline 

0133+476	&	C	&	0.86	&	5337	&	14.4	&	2.5	&	20.7	&$(	0.9	\pm	1.3	)\times10^{10}$&$	0.3	\pm	0.3	$\\
0138-097	&	C	&	0.73	&	4381	&	-	&	-	&	-	&$(	2.3	\pm	1.0	)\times10^{10}$&$		-		$\\
0256+075	&	A	&	0.89	&	5602	&	-	&	-	&	-	&$(	5.4	\pm	1.2	)\times10^{10}$&$		-		$\\
0420-014	&	B	&	0.92	&	5770	&	11.4	&	1.9	&	19.9	&$(	6.2	\pm	1.1	)\times10^{10}$&$	1.7	\pm	0.3	$\\
0745+241	&	B	&	0.41	&	2164	&	-	&	-	&	-	&$(	6.2	\pm	1.2	)\times10^{10}$&$		-		$\\
0804+499	&	A	&	1.43	&	10427	&	17.8	&	0.2	&	35.5	&$(	1.0	\pm	1.2	)\times10^{10}$&$	1.2	\pm	0.9	$\\
0814+425	&	C	&	0.25	&	2952	&	2.7	&	8.4	&	4.6	&$(	1.0	\pm	1.1	)\times10^{10}$&$	0.4	\pm	0.2	$\\
0859+470	&	B	&	1.46	&	10455	&	-	&	-	&	-	&$(	1.0	\pm	1.1	)\times10^{10}$&$		-		$\\
0906+430	&	A	&	0.67	&	3934	&	-	&	-	&	-	&$(	3.7	\pm	1.3	)\times10^{10}$&$		-		$\\
0953+254	&	B	&	0.71	&	4208	&	-	&	-	&	4.3	&$(	3.6	\pm	1.0	)\times10^{10}$&$		-		$\\
1055+018	&	B	&	0.89	&	5593	&	11.1	&	4.7	&	12.2	&$	-					$&$		-		$\\
1156+295	&	A	&	0.73	&	4332	&	25.1	&	2.0	&	28.5	&$(	2.1	\pm	1.1	)\times10^{10}$&$	0.6	\pm	0.2	$\\
1510-089	&	B	&	0.36	&	1860	&	20.7	&	3.4	&	16.7	&$(	1.8	\pm	0.9	)\times10^{10}$&$	0.5	\pm	0.2	$\\
1611+343	&	C	&	1.40	&	9809	&	14.2	&	4.2	&	13.7	&$(	2.8	\pm	0.9	)\times10^{10}$&$	0.9	\pm	0.3	$\\
1633+382	&	A	&	1.81	&	13569	&	30.5	&	2.5	&	21.5	&$	-					$&$		-		$\\
1637+574	&	B	&	0.75	&	4519	&	11.0	&	4.0	&	14.0	&$(	2.9	\pm	1.3	)\times10^{10}$&$	0.8	\pm	0.2	$\\
1641+399	&	A	&	0.59	&	13569	&	27.7	&	5.1	&	7.8	&$(	0.9	\pm	3.9	)\times10^{10}$&$	0.4	\pm	0.3	$\\
1652+398	&	C	&	0.03	&	142	&	-	&	-	&	-	&$(	0.4	\pm	1.2	)\times10^{10}$&$		-		$\\
1739+522	&	B	&	1.38	&	9612	&	-	&	-	&	26.5	&$(	1.2	\pm	1.2	)\times10^{10}$&$		-		$\\
1954+513	&	B	&	1.22	&	8274	&	-	&	-	&	7.4	&$(	4.4	\pm	1.1	)\times10^{10}$&$		-		$\\
2134+004	&	A	&	1.93	&	14664	&	9.0	&	2.2	&	16.1	&$(	6.7	\pm	1.1	)\times10^{10}$&$	2.3	\pm	0.6	$\\
2145+067	&	C	&	0.99	&	6368	&	8.0	&	1.1	&	15.6	&$(	1.9	\pm	1.2	)\times10^{10}$&$	1.1	\pm	0.6	$\\
2230+114	&	A	&	1.04	&	6747	&	15.5	&	3.7	&	15.6	&$(	1.3	\pm	0.8	)\times10^{10}$&$	0.5	\pm	0.2	$\\
2251-152	&	C	&	0.86	&	5333	&	19.9	&	1.3	&	33.2	&$(	1.1	\pm	1.4	)\times10^{10}$&$	0.4	\pm	0.3	$\\
3C 279		&	B	&	0.54	&	2998	&	20.9	&	2.4	&	24.0	&$(	2.7	\pm	0.1	)\times10^{10}$&$	0.7	\pm	0.2	$\\
OJ 287		&	C	&	0.31	&	1540	&	15.4	&	3.3	&	17.0	&$(	0.5	\pm	0.2	)\times10^{10}$&$	0.2	\pm	0.1	$\\

\hline 
\multicolumn{9}{l}{$^\dagger$ from Hovatta et al (2009)}
\end{tabular}\\
\end{center}
\label{tab:Bfield3freqs}
\end{table*}

The magnetic fields obtained range from $\sim0.2$ Gauss (OJ287, epoch C) to $\sim2.3$ Gauss (2134+004, epoch A), with the typical magnetic field strengths being of the order of half a Gauss. Comparing the inferred magnetic fields for the two sources available for both epochs reveals that 
$B_{1}$ is the same within the uncertainties for 0133+476, but differs by about a factor of two for 0420--014 (this difference is $\sim2.1\sigma$) between our two epochs. 

\begin{figure}
\begin{center}
\includegraphics[width=8cm]{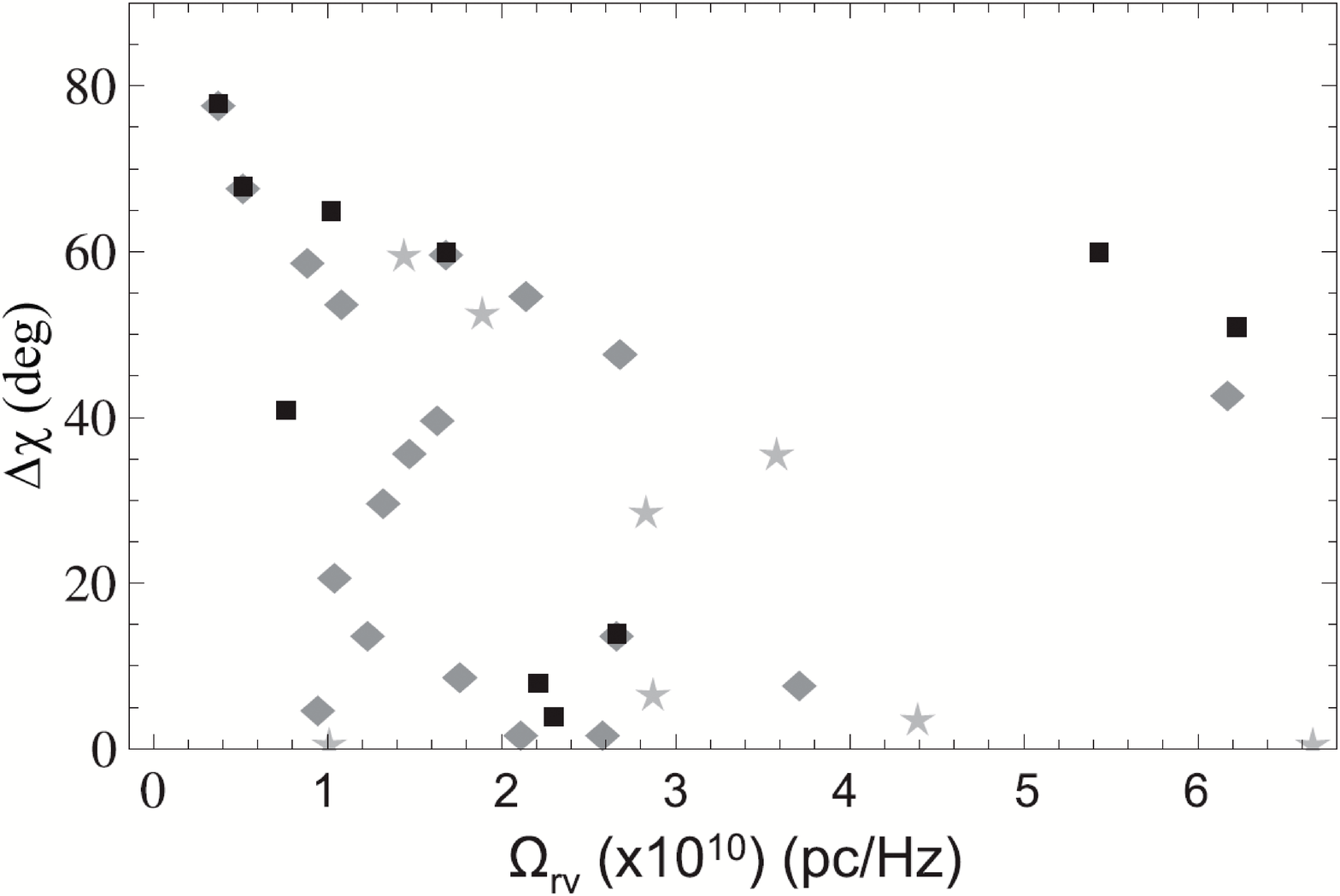}\hspace{0.8cm}\\ \vspace{0.5cm}
\includegraphics[width=8cm]{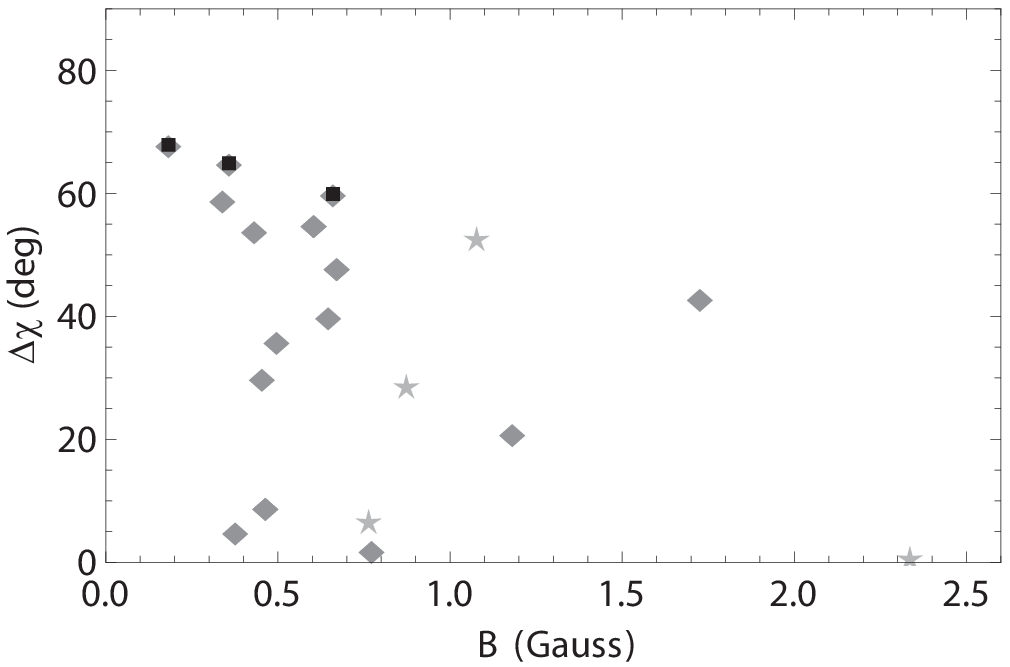}\\
\end{center}
\caption{Core shift parameter $\Omega_{r\nu}$ (upper) and magnetic field at 1~pc (lower) plotted against $\Delta\chi$. Symbols have the same significance as in Figure 2. }
\label{fig:VS_Bfield}
\end{figure}

Comparisons of the magnetic field strength with $\Delta\chi$ are shown in Fig.~\ref{fig:VS_Bfield}. 
The correlation coefficients are found to be  --0.17 for $\Omega_{r\nu}$ and --0.38 for $B_1$. However, the maximum observed $\Delta\chi$ tends to decrease as the magnetic field increases. If this is true, this suggests that large misalignments in $\Delta\chi$ could be associated in part with relatively low core magnetic fields.

A plausible explanation is that as the magnetic field increases, any turbulence, random component or disruption in the AGN would be of less relative importance. Under those circumstances, it is reasonable to suppose that optical and radio properties might become more and more correlated and thus, $\Delta\chi$ will tend to be small. 
Another way to look at this hypothesis is that the jet magnetic fields are not sufficiently well ordered over sufficiently large regions to ensure that $\Delta\chi$ is always small.

\section{Conclusions}

Our results in Paper I indicated that the difference between the VLBA radio core and optical polarizations of AGNs showed a more complex behaviour than was expected based on the results of Gabuzda et. al (2006). We have considered here simultaneous observations at 5 frequencies (12+15+22+24+43~GHz) and in the optical, in order to determine if effects such as internal or high external Faraday rotation could be affecting our radio data.

With these new data, we have been able to resolve some ambiguities in the rotation measure, and confirm the presence of relatively high Faraday rotation in some of the observed cores. However, no extremely high rotation measures were found, and the values derived from our five-frequency data are generally consistent with previous estimates based on three frequencies.

The behaviour of the polarization angles can be complex. Some deviations from $\lambda^2$ behaviour are found at high frequencies (usually for 43~GHz), restricting the frequency range that can effectively be used to derive the core RMs. It is possible that these deviations arise because different regions in the source or different source structures are being probed at different frequencies.

Using five-frequency data to estimate the zero-wavelength radio core polarization angles for the eight sources considered here and comparing these with the simultaneously measured optical polarization angles yields no clear peaks in the distribution of $\Delta\chi$; in particular, no clear peak at $\Delta\chi\sim 0$ has emerged for the 8 AGN considered here. We have attempted to identify some characteristic property of the cores for which $\Delta\chi\sim0$, but without success. 
We have extended the analysis in Paper I by examining the data correlations between $\Delta\chi$ and the core rotation measures and the magnetic fields at 1~pc from the jet base. We have found no evidence for correlations between any of these properties and the observed $\Delta\chi$ values. Thus, although the overall distribution of $\Delta\chi$ for all 40 sources in our sample does show a significant peak at $\Delta\chi\sim 0$ (Paper~I), it remains unclear what distinguishes these AGN cores from those having misaligned polarization position angles.

Thus, $\Delta\chi\sim0$ for some AGN, consistent with co-spatial emission in the optical and radio, Misaligned polarization position angles possibly indicate 
that the optical and radio emission in those objects arises in different parts of the jet with different magnetic field orientations. There are cases where $\Delta\chi$  remains constant for a particular source, but there are also counterexamples of the optical and VLBI core polarization angles changing with time. 
We do not currently have sufficient information to identify the origin of this behaviour, and further coordinated optical and VLBA observations would be of interest in this connection.

We have estimated the core-shift exponents $k_r$ for the eight AGN considered here. The fitted $k_r$ values for six of the eight AGN are equal to unity within the uncertainties, supporting the previous conclusion of O'Sullivan \& Gabuzda (2009)  that the core regions of radio-loud AGN are often near equipartition. The derived core magnetic fields at 1~pc from the jet base for four AGNs are of the order of half a Gauss, similar to the values found for other sources by O'Sullivan \& Gabuzda (2009), based on VLBA data obtained simultaneously at seven frequencies.

We have found no evidence for clear correlations between any of these properties and the observed $\Delta\chi$ values, although there is a hint that the maximum observed $\Delta\chi$ may tend to decrease as the core-region magnetic field increases, suggesting that large misalignments in $\Delta\chi$ could be more common in cores with relatively low magnetic  fields.

\section*{Acknowledgements}
This publication has emanated from research conducted with the financial support of Science Foundation Ireland. P.S.S acknowledges support from NASA/JPL/Spitzer contract 1256424 and NASA/Fermi Guest Investigator grants NNX08AW56G and NNX09AU10G. The National Radio Astronomy Observatory is operated by Associated Universities Inc. We thank the referee Dr. Philip Hughes for his prompt reading of the manuscript and helpful suggestions for revisions.

\label{lastpage}


\begin{thebibliography}{99}
\bibitem[]{Algaba}Algaba, J. C., 2010, PhD Thesis, University College Cork
\bibitem[]{Algabaetal}Algaba, J. C., Gabuzda, D. C., Smith, P.S., 2010, MNRAS 411, 85
\bibitem[]{D'Arcangelo07}D'Arcangelo, F. D., Marscher, A. P., Jorstad, S. G.,  Smith, P. S., Larionov, V. M., Hagen-Thorn, V. A., Kopatskaya, E. N.,  Williams, G. G., Gear, W. K., 2007, ApJ, 659, L107
\bibitem[]{Gabuzda03}Gabuzda, D. C. 2003, ApSS, 288, 39
\bibitem[]{Gabuzda06}Gabuzda, D. C., Rastorgueva, E. A., Smith, P. A., O'Sullivan, S. P., 2006,  MNRAS 369,1596
\bibitem[Ghisellini (1985)]{Ghisellini}Ghisellini, G., Maraschi, L., Treves, A. 1985, A\&A, 146, 204
\bibitem[]{VLBAmem30}G\'omez, J. L., Marscher, A. P., Alberdi, A, Jorstad, S., Agudo, I. 1992, NRAO VLBA Memo 30
\bibitem[]{Hovatta09}Hovatta, T.; Valtaoja, E.; Tornikoski, M.; L\"ahteenm\"aki, A. (2009) A\&A, 494, 527
\bibitem[]{Jorstad07}Jorstad, S. G., Marscher, A. P., Stevens, J. A., Smith, P. S., Forster, J. R., Gear, W. K., Cawthorne, T. V., Lister, M. L., Stirling, A. M., G\'omez, J. L., Greaves, J. S., Robson, E. I. 2007, AJ, 134, 799
\bibitem[]{Konigl81}K\"onigl A., 1981, ApJ, 243, 700
\bibitem[]{Lobanov98}Lobanov, A. P., 1998 A\&A 330, 79
\bibitem[]{Mahmud08}Mahmud, M.; Gabuzda, D. C. Proceedings for `The 9th European VLBI Network Symposium on The role of VLBI in the Golden Age for Radio Astronomy and EVN Users Meeting' September 23-26th 2008, Bologna, Italy
\bibitem[]{Mahmud09}Mahmud, M.; Gabuzda, D. C.; Bezrukovs, V., 2009, MNRAS, 400, 2
\bibitem[]{OSullivan09b}O' Sullivan, S. P., Gabuzda, D. C., 2009, MNRAS, 400, 26
\bibitem[]{Pradel06}Pradel, N.; Charlot, P.; Lestrade, J.-F., 2006, A\&A, 452, 1099
\bibitem[]{Schmidt92a}Schmidt, G. D., Stockman, H. S., Smith, P. S., 1992a, ApJ 398, L57
\bibitem[]{Schmidt92b}Schmidt, G. D., Elston, R. \& Lupie, O. L., 1992b, AJ 104, 1563
\bibitem[]{Smith03}Smith, P. S., Schmidt, G. D., Hines, D. C., Foltz, C. B., 2003, ApJ, 593, 676
\bibitem[]{WardleKronberg}Wardle, J. F. C, Kronberg, P. P., 1974, ApJ, 194, 249
\bibitem[]{ZavalaTaylor01}Zavala R. T., Taylor G. B., 2001, ApJ, 550, 147
\end{thebibliography}
\end{document}